\documentstyle[epsfig]{mn}
\def\figdir{.}

\def\epsscale#1{\epsfxsize=#1\columnwidth}
\def\plotone#1{\par\centerline{\epsfbox{#1}}}

\def\etal{{ et al.~\/}}

\title[Morphology and Evolution of Galaxy Clusters]
{Morphology and Evolution of Simulated and Optical Clusters: 
A Comparative Analysis} 

\author[Rahman, Krywult, Motl, Flin, and Shandarin]
{Nurur Rahman$^1$
\thanks{National Research Council (NRC) postdoctoral fellow; 
Current address: IPAC/California Institute of Technology, 
Mail Code 100-22, Pasadena, CA 91125, USA; 
Email: nurur@ipac.caltech.edu}, 
Janusz Krywult$^2$, 
Patrick M. Motl$^3$,
Piotr Flin$^2$, and 
Sergei F. Shandarin$^1$ \\ 
$^1$Department of Physics and Astronomy,
University of Kansas, Lawrence, KS 66045, USA;\\
$^2$Institue of Physics, Pedagogical University, 
Kielce, Poland; \\
$^3$Department of Physics an Astronomy, 
Louisiana State University, Baton Rouge, LA 70803, USA;\\
nurur@kusmos.phsx.ku.edu, krywult@pu.kielce.pl, 
motl@rouge.phys.lsu.edu, sfflin@cyf-kr.edu.pl, sergei@ku.edu
}
\date{}
  
\begin{document}
\maketitle

\begin{abstract}
We have made a comparative study of morphological evolution in simulated 
DM halos and X-ray brightness distribution, and in optical clusters. Samples 
of simulated clusters include star formation 
with supernovae feedback, radiative cooling, and simulation in the adiabatic 
limit at three different redshifts, $z = 0.0, 0.10,$ and $0.25$. 
The optical sample contains 208 ACO clusters within redshift, $z \leq 0.25$.
Cluster morphology, within 0.5 and 1.0 h$^{-1}$ Mpc from cluster center, is 
quantified by multiplicity and ellipticity. 

We find that the distribution of the dark matter halos in the adiabatic 
simulation appear to be more elongated than the galaxy clusters. Radiative 
cooling brings halo shapes in excellent agreement with observed clusters, 
however, cooling along with feedback mechanism make the halos more flattened.

Our results indicate relatively stronger structural evolution and more 
clumpy distributions in observed clusters than in the structure of simulated 
clusters, and slower increase in simulated cluster shapes compared to those 
in the observed one. 

Within $z \leq 0.1$, we notice an interesting agreement in the shapes of 
clusters obtained from the cooling simulations and observation. We also 
notice that the different samples of observed clusters differ significantly 
in morphological evolution with redshift. We highlight a few possibilities 
responsible for the discrepancy in morphological evolution of simulated and 
observed clusters. 
\end{abstract}
\section{Introduction}
The hierarchical clustering is the most popular model for the Large Scale 
Structure (LSS) formation. The model relies on the assumption that larger 
structures result from the merging of smaller sub-clumps. 
Theoretical paradigm of the hierarchical evolution is the Cold Dark Matter 
(CDM) scenario which assumes that baryonic matter (stars, hot X-ray 
gas) evolves in the dark matter (DM) potential through violent processes. 
Structural evolution in cosmological objects such as galaxies or clusters 
of galaxies is the underlying principle in this scenario. 
A generic prediction of the CDM model is the nonsphericity of the DM halos.  
The degree of flattening of the halos evolves in cosmological time, from 
highly irregular at the distant past towards more regular at the present. 
In principle, the model prediction can be tested comparing the DM halo shapes 
with that of the (baryonic) matter distributions. A comparative morphological 
analysis between model and observation could help constraining the nature of 
the DM and its role on the LSS.

Melott, Chambers \& Miller (2001; hereafter MCM) have reported evolution in 
the gross morphology of galaxy-clusters (quantified by ellipticity) for a 
variety of optical and X-ray samples for $z < 0.1$. They infer that the 
evidence is consistent with a low matter density universe. 
Using a similar shape measure as well as intra-cluster medium temperature and 
X-ray luminosity, Plionis (2002) has presented evidence for recent evolution 
in optical and X-ray cluster of galaxies for $z \leq 0.18$. In both studies 
evolution is quantified by the change of cluster ellipticity with redshift. 
In a recent study, Jeltema et al. (2005) have reported structural evolution 
of clusters with redshift where cluster morphology is quantified by the power 
ratio method (Buote \& Tsai 1995). Jeltema et al. used a sample of 40 X-ray 
clusters over the redshift range $\sim 0.1 - 0.8$ obtained from Chandra 
Observatory. 
In spite of methodological differences, the results of these studies indicate 
evolution in the morphology of the largest gravitationally bound systems over 
a wide range of look-back time. 

The observational evidence prompted concerns about the formation and evolution 
of structures in the CDM scenario via numerical simulations. If the results of 
simulations provide faithful representations of the evolutionary history of 
cosmological objects, then one would expect a similar trend in the structure 
of simulated objects. So far almost all studies of simulated clusters are  
focused either on understanding the nature of the background 
cosmology within which the present universe is evolving (Jing et al. 1995; 
Crone, Evrard \& Richstone 1996; Buote \& Xu 1997; Thomas et al. 1998; 
Valdarnini, Ghizzardi \& Bonometto 1999; Suwa et al. 2003) or on understanding 
the distribution and shape of the DM halos in various types of simulations, 
e.g., simulations with or without baryons and gas physics (Dubinski \& Carlberg 
1991; Dubinski 1994; Aninos \& Norman 1996; 
Tissera \& Dominguez-Tenreiro 1998; Bullock 2001; Buote et al. 2002; Jung \& 
Suto 2002; Gao et al. 2004a,b; Kazantzidis et al. 2004; Springel, 
White \& Hernquist 2004; Allgood et al. 2005; Nagai \& Kravtsov 2005; 
Flores et al. 2005; Libeskind et al. 2005; Maccio et al. 2005; van der Bosch 
et al. 2005; Zentner et al. 2005). Until recently a comparative study of 
morphological evolution in simulated and real clusters was absent. Floor et 
al. (2003) and Floor, Melott \& Motl (2004; hereafter FMM) have investigated 
evolution in cluster morphology simulated with different initial conditions, 
background cosmology, and different physics (e.g. simulation with or without 
radiative cooling). They have used eccentricity as a probe to quantify 
evolution. Their studies, emphasizing shape in the outer regions of clusters, 
suggest slow evolution in simulated cluster shapes compared to the observed 
one. However, the studies of Floor and collaborators are indirect in a sense 
that they did not analyze observed clusters using the same measurement 
technique applied to their simulated data sets.

In this paper we make a comparative analysis between simulated and observed 
clusters where both data sets are juxtaposed and analyzed using the same set 
of structural measures. We analyze cluster morphology and its evolution using 
shape measures such as multiplicity ($M$) and ellipticity ($\epsilon$) derived 
from the Minkowski functionals (Rahman \& Shandarin 2003, 2004, hereafter RS03 
and RS04; Rahman et al. 2004). 
The MFs provide a non-parametric description of the images with no prior 
assumptions made on the shapes of the images. The measurements based on the 
MFs appear to be robust and numerically efficient when applied to various 
cosmological studies, e.g., galaxies, galaxy-clusters, CMB maps etc. (Mecke, 
Buchert \& Wagner 1994; Schmalzing et al. 1999; Beisbart 2000; Beisbart, 
Buchert \& Wagner 2001; Beisbart, Valdarnini \& Buchert 2001; Kerscher et al. 
2001a, 2001b; Shandarin, Sheth \& Sahni 2004). Various measures, constructed 
from the two-dimensional scalar, vector, and several tensor MFs have been 
described and tested in RS03 and RS04. To derive the parameters applied in 
this study we use the extended version of the numerical code developed in RS03 
and RS04. 

We study evolution in the simulated clusters in a flat CDM universe 
($\Lambda$CDM; $\Omega_m = 0.3$, $\Omega_{\Lambda} = 0.7$) obtained from 
three different sets of high resolution simulations (Motl et al. 2004). 
The first set has clusters simulated in the adiabatic limit, the second set 
contains clusters with radiative cooling (RC), and the last set includes 
clusters with cooling + star formation and supernovae feedback (SFF). Each 
sample contain DM as well as X-ray brightness distributions at three 
different redshifts, $z = 0.0, 0.1,$ and $0.25,$. For comparision we also 
analyze a sample of ACO (Abell, Corwin \& Olowin 1989) clusters within 
$z \leq 0.25$. The sample contains 208 optical clusters derived from 
10-inch photographic plates taken with the 48-inch Palomar Schmidt 
Telescope (Tr\`{e}vese et al. 1992; Flin et al. 1995; Tr\`{e}vese et al. 
1997; Flin et al. 2000). 

The objective of our study is twofold: first, to check the efficiency of 
the parameters differentiating various sets of objects, and second, to 
explore (statistical) correspondence in the morphological properties of the 
distributions of DM halos, X-ray emitting gas, and optical clusters using 
measures that are sensitive to shape and sub-structures.

In the CDM model (satellite) galaxies are associated with the DM sub-halos 
that are accreted by their (current) parent halo, a bigger structure 
usually associated with a galaxy cluster. If this is the case statistical 
properties of galaxies regarding mass, sub-structure, shape etc., would 
show a similar trend to that of the sub-halos. 
On the other hand X-ray emitting hot gas, evolving in the DM background 
potential, would not directly follow the DM distribution because of its 
isotropic pressure support. Therefore, a statistical analysis of various 
properties of DM halos, galaxy clusters, and X-ray gas distributions will 
be useful to probe possible bias of luminous galaxies toward sub-halos 
and their correspondence with the distribution of hot gas. This is the 
motivation behind the second objective.

The organization of the paper is as follows: simulation technique and the 
observational data are described briefly in $\S2$, a short discussion of 
shape measures is given in $\S3$. The results are presented in $\S4$ and 
the conclusions are summarized in $\S5$.
\section{Data}
\subsection{Numerical Simulations}
We have analyzed images of simulated clusters projected along three 
orthogonal axes. The clusters have been simulated in the standard, flat 
cold DM universe ($\Lambda$CDM) with the following parameters: 
$\Omega_b = 0.026$, $\Omega_m = 0.3$, $\Omega_{\Lambda} = 0.7$, $h = 0.7$, 
and $\sigma_8 = 0.928$. For a complete description of the simulations see 
Motl et al. (2004). We have used three samples of clusters derived from 
the same initial conditions and background cosmology. The difference 
between the samples is in the energy loss mechanism experienced by the 
baryonic fluids. In the first sample, no energy lose is allowed; in the 
second sample fluid is allowed to lose energy via radiation and 
subsequently cool; in the third sample, physics of star formation and 
supernovae feedback are incorporated in addition to radiative cooling.

The simulations use a coupled N-body Eulerian hydrodynamics code (Norman 
\& Bryan 1999; Bryan, Abel \& Norman 2000) where the dark matter particles 
are evolved by an adaptive particle-mesh, N-body code.  
The PPM scheme (Colella \& Woodward 1984) is used to treat the fluid 
component on a comoving grid. An adaptive mesh refinement (AMR) is employed 
to concentrate the numerical resolution on the collapsed structures that 
form naturally in cosmological simulations. 
The DM particles exist on the coarsest three grids; each sub-grid having 
twice the spatial resolution in each dimension and eight times the mass 
resolution relative to its parent grid. At the finest level, each particle 
has a mass of $9 \times 10^{9} \; \mathrm{h}^{-1} \; \mathrm{M_\odot}$.
A second order accurate TSC interpolation is used for the adaptive particle 
mesh algorithm.  Up to seven levels of refinement are utilized for the 
fluid component, yielding a peak resolution of $15.6 \: \mathrm{h}^{-1}$ 
kpc within the simulation box with sides of length 256 $\mathrm{h}^{-1}$ 
Mpc at the present epoch. Clusters are selected using the HOP algorithm 
(Eisenstein \& Hut, 1998) with an overdensity threshold of 160. 

A tabulated cooling curve (Westbury \& Henriksen 1992) for a plasma of 
fixed, 0.3 solar abundance has been used to determine the energy loss to 
radiation. Heat transport by conduction is neglected in the present 
simulations since it has been shown that even a weak, ordered magnetic 
field can reduce conduction by two to three orders of magnitude from the 
Spitzer value (Chandran \& Cowley 1998). However, Narayan \& Medvedev 
(2001) has shown that if the chaotic magnetic field fluctuations extends 
over a sufficiently large length scales within the intra-cluster medium 
(ICM), then thermal conductivity becomes significant to the global energy 
balance of the ICM. Energy input into the fluid from AGN is also neglected 
in the current simulations. 

The prescription of Cen \& Ostriker (1992) has been used to transform 
collapsing and rapidly cooling gas into collisionless star particles. At 
the finest resolution level, a grid cell is eligible to form a star in a 
given time step if the local flow is converging, the dynamical time 
exceeds the cooling time and a Jean's mass worth of gas exists within the 
cell.To model a population of prompt supernovae, thermal feedback has been 
introduced. The amount of feedback has been set from numerical experiments 
to provide a reasonable amount of mass in star particles. The feedback is 
approximately $7 \times 10^{48}$ ergs per solar mass of stars formed or 
about half a keV of energy per particle in the final clusters. 

We have 41 three dimensional clusters from each sample, giving a total of 123 
projected clusters in the respective samples. Each projection is constructed 
within a 8 h$^{-1}$ Mpc (comoving) frame containing 360 $\times$ 360 pixels. 
Majority of clusters in each sample is in the mass range $\sim 10^{13}-10^{14}$ 
M$_{\odot}$ with few clusters ($\sim 15$) in the limit $\sim 10^{15}$ M$_{\odot}$. 
\subsection{Optical Clusters}
The details of data acquisition and processing of the optical sample has been 
described in Tr\`{e}vese et al. 1992; Flin et al. 1995; Tr\`{e}vese et al. 
1997; Flin et al. 2000. Here we highlight only the essential features of the 
sample needed for this study. 
  
The sample contains 208 optical clusters, within $z \leq 0.25$, derived from 
10-inch photographic plates taken with the 48-inch Palomar Schmidt Telescope. 
It contains rich and massive ACO clusters with richness $R \geq 1$ and mass, 
approximately, in the range $\sim 10^{13}-10^{14}$ M$_{\odot}$. Highly  
massive structures, e.g. Coma cluster (A1654) or clusters constituting Shapley 
condensation, are absent in this sample.

The visual control is the greatest advantage of this sample. The essential 
difference between this and other samples is visual control of all objects 
classified as galaxies when automatic procedure were applied. The visual 
inspection was made for objects with magnitude range at least $m_3 + 3$ mag. 
The relationship between the number of objects with respect to the magnitude 
and the luminosity function for each separate cluster show that clusters 
are complete at least in the magnitude range $m_3$ to $m_{2.5}$. In majority 
of cases, it is complete till $m_3 + 3$. 
\section{Morphological Parameters}
We use multiplicity ($M$) and ellipticity ($\epsilon$) as quantitative measures 
to study evolution of observed and simulated clusters. Ellipticity is derived 
from the area tensor functional, a member of the hierarchical set of the MFs. 
This functional is given by, 
\begin{equation}
A_{ij} = \int\limits_{K} (x_i-A_i) (x_j-A_j) \ da, 
\label{tA}
\end{equation}
where $K$ is the region bounded by a given contour and $A_i$ is the area 
vector functional, i.e. area centroid, expressed as follows, 
\begin{equation}
A_i   = \frac{1}{A_S}\int\limits_{K} x_i \ da. 
\label{vA} 
\end{equation}
The symbol $A_S$ represents the area within the contour. It is known as the 
scalar area functional and is given by, 
\begin{equation}
A = \int\limits_{K} \ da. 
\label{sA} 
\end{equation}
The area vector functional is in fact the center of mass of the region 
within the contour if we assume that the surface density of the (enclosed) 
region is constant. The tensor $A_{ij}$ is closely related to the inertia 
tensor of a homogeneous region, The details of the MFs can be found in 
Schmalzing (1999), Beisbart (2000), and RS03. 

$\bullet$ Multiplicity ($M$): This parameter is defined as, 
\begin{equation}
M = \frac{1}{A_{max}} \ \sum_{i=1}^N A_i = \frac{A_S}{A_{max}},
\label{mul}
\end{equation}
where $A_i$ is area of the individual components at a given level, $A_{max}$ 
is the area of the largest component at that level , $N$ is the total number 
of components, and $A_S$ is the total area at that level obtained after 
summing the areas of the components. Multiplicity, $M = M (A_S)$, is a measure 
with fractional value and gives the number of components measured at any 
brightness level: $M = 1$ for a single iso-intensity contour, i.e. component, 
and $M > 1$ for multi-contours. 
 
It may be mentioned here that Thomas et al. (1998) have used multiplicity 
as a parameter for sub-structure measure in N-body simulations. They define 
it as a ratio of mass of sub-clumps to cluster mass. In this study it is a 
ratio of the areas (sizes) as defined in equation 1. 

We use two variants of $M$ to present our results: one is the average of 
multiplicity over all density/brightness levels, $\bar{M}_{eff}$, and 
the other is the maximum of the multiplicity found at one of the levels, 
$M_{max}$.

$\bullet$ Ellipticity ($\epsilon$): We adopt the definition of ellipticity, 
\begin{equation} 
\epsilon = 1 - b/a,
\label{ell1}
\end{equation} 
where $a$ and $b$ are the semi-axes of an ellipse. For our purpose the 
semi-axes correspond to the ``auxiliary ellipse'' constructed from the 
eigenvalues of the area tensor (see RS03 for detail). Note that the 
``auxiliary ellipse'' is an ellipse having exactly the same area tensor.
  
We have used two variants of $\epsilon$: one is sensitive to the shape 
of the individual cluster components present at a given level while the 
other is sensitive to the collective shape formed by all the components 
present at that level. 
We label these two variants of $\epsilon$, respectively, as the effective 
(${\epsilon}_{eff}$) and the aggregate (${\epsilon}_{agg}$) ellipticity. 
Morphological properties of clusters such as shape and the nature or the 
degree of irregularity existing in these systems can be probed effectively 
with these two parameters.

At any given density/brightness level, we construct ${\epsilon}_{eff}$ 
as a weighted mean normalized by the multiplicity and area of the largest 
contour, 
\begin{equation} 
{\epsilon}_{eff} = \frac{1}{M \cdot \ A_{max}} \ 
\sum_{i=1}^N \epsilon_i \ A_i,
\label{ell2}
\end{equation} 
where $\epsilon_i$ are ellipticities of the individual components measured 
as stated earlier and $M$ is the multiplicity at that level. The symbols 
$A_i$ and $A_{max}$ have similar meanings as before. This measure can be 
used as an effective tool to quantify shapes of large scale merger remnants.

To construct ${\epsilon}_{agg}$, we take the union of all components present 
at a given level and form a collective region. The integrated region can be 
expressed as 
\begin{equation} 
R = R_1 \cup R_2 \cup \cdot \cdot \cdot \cup R_N, 
\label{ell3}
\end{equation} 
where $R_i$ is the region enclosed by each contour. Subsequently we find 
the components of the area tensor and the ``auxiliary ellipse'' for the 
region $R$. 

The behavior of ${\epsilon}_{agg}$ is similar to the conventional 
ellipticity measure based on the inertia tensor (Carter \& Metcalf 1980). 
But the construction procedure of these two measures are different. 
The conventional method finds the eigenvalue of the inertia tensor for an 
annular region enclosing mass density or surface brightness. On the other 
hand, the method based on MFs finds the eigenvalues of regions enclosed by 
the contour(s) where the regions are assumed to be homogeneous.   

We have computed ellipticities after averaging the estimates at all 
density/brightness levels. Our final result is, therefore, expressed as  
$\bar{\epsilon}_{eff}$ and $\bar{\epsilon}_{agg}$. 
\subsection{Toy Models}
To get a better feeling of the parameters mentioned above we provide an 
illustrative example with toy models. We find this demonstration useful 
since it gives a visual expression how the number of group members forming 
complex structures affects the shapes (see also Paz et al. 2005).  

\begin{figure}
\epsscale{1.10}
\plotone{\figdir/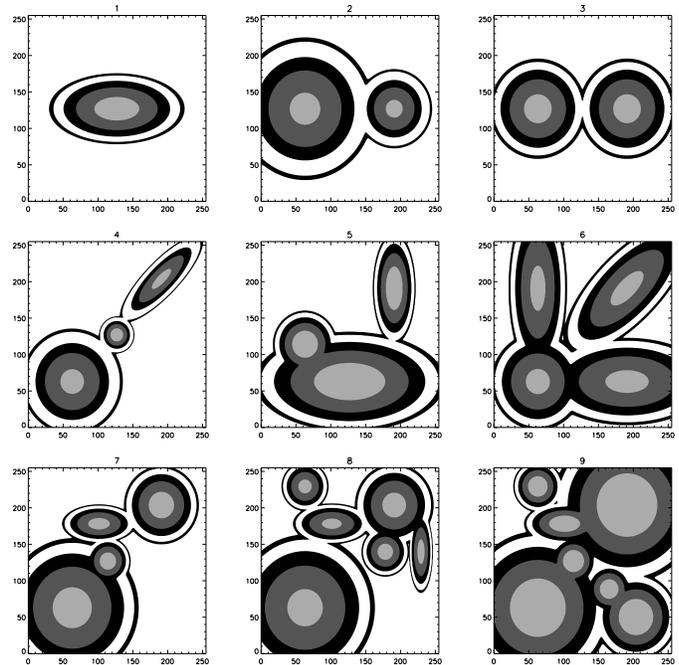}
\caption{Contour plots of toy clusters at different brightness levels 
(in arbitrary scales). The multi-modal clusters have clumps with 
different peak brightness. For all clusters the outer line represents 
the percolation level where the sub-structures merge and form a single, 
large system.
\label{toy_img}}
\end{figure}

One can think of these toy images as snap shots of different clusters (in 
projection) taken at one particular time. We include clusters with different 
types of internal structures in Fig. \ref{toy_img}: unimodal elliptic 
structure (panel 1), asymmetric and symmetric bimodal clusters (panels 2 
and 3, respectively), cluster with filamentary structure (panel 4) etc. 
The multi-modal clusters have clumps with different peak brightness. We 
show contour plots of toy clusters at different brightness levels where 
the levels are chosen arbitrarily. For all clusters the outer line 
represents the percolation level where the sub-structures merge with one 
another and form a single, large system. 

\begin{figure}
\epsscale{1.10}
\plotone{\figdir/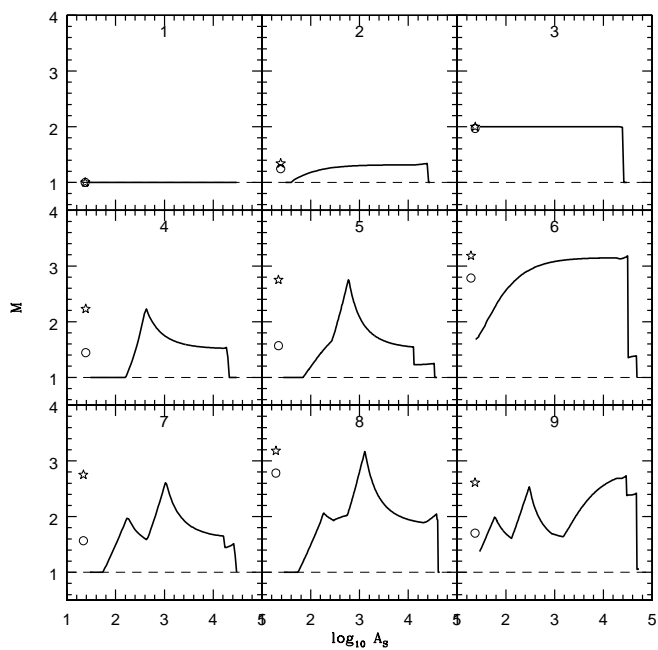}
\caption{Multiplicity as a function of contour area ($A_S$) for toy clusters  
as shown in Fig. \ref{toy_img}. The circle (star) represents the effective 
multiplicity $\bar{M}_{eff}$ (maximum multiplicity $M_{max}$) as defined in 
the text. The x-coordinates of these legends are chosen only for the 
convenience of demonstration. Recall that the position of the highest peak 
along the x-axis corresponds to the $M_{max}$ whereas $\bar{M}_{eff}$ is 
obtained after averaging along the x-axis. See text for details. 
\label{toy_mul}}
\end{figure}

Multiplicity as a function of component area (in grid units) is shown in 
Fig. \ref{toy_mul} for our selected toy models. As mentioned earlier $M$ 
is sensitive to the size of the sub-structures. The simplest case to see 
this is a bimodal cluster. 
For a bimodal structure with unequal sub-clumps (panel 2), the fractional 
value of multiplicity ($1 < M < 2$) tells us that the components of the 
system have different sizes. The isolated components eventually percolate 
giving $M = 1$ at low brightness level, i.e, at larger area. 
On the other hand, for a cluster with equal components $M = 2$ until 
percolation occurs (panel 3). For clusters with three components 
(panels 4 and 5), we see that for a small range of brightness levels, 
the components are well separated where two of these are bigger then the 
third one ($2 < M < 3$). Afterward two of the three clumps merge together 
giving $1 < M < 2$. These two remaining components eventually 
percolate to become a single system. The clumps in panel 6 are distributed 
around the center. For this cluster, we see two unequal but well separated 
clumps ($1 < M < 2$) with the same peak brightness. The behavior of clusters 
in panels 7 and 8 is similar except that they have a different number of 
sub-structures. The cluster in panel 9 has the largest number of components 
(a total of 7). Two of its clumps are so large compared to the other ones 
that they dominant. The multiplicity is always in the range $1 < M < 3$ 
reflecting the merger of clumps at different levels. 

\begin{figure}
\epsscale{1.10}
\plotone{\figdir/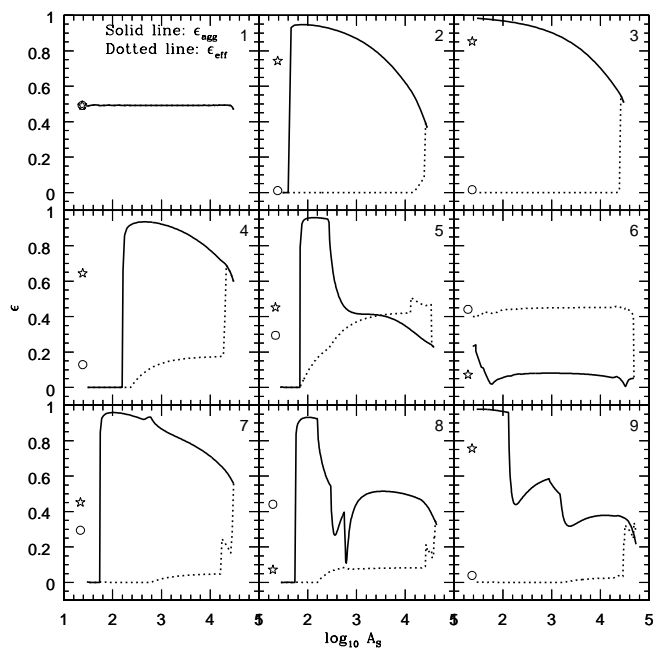}
\caption{Ellipticity as a function of contour area ($A_S$) for toy models 
as shown in Fig. \ref{toy_img}. Dotted and solid lines represent, 
respectively, $\epsilon_{eff}$ and $\epsilon_{agg}$. The circle (star) 
represents the $\bar{e}_{eff}$ ($\bar{\epsilon}_{agg}$) for these 
toy clusters as defined in the text. Once again the x-coordinates of these 
legends are chosen only for the convenience of demonstration. See text for 
details. \label{toy_ell}}
\end{figure}

Ellipticity for these toy clusters is shown in  Fig. \ref{toy_ell}. In this 
figure the solid and dotted line represent, respectively, $\epsilon_{agg}$ 
and $\epsilon_{eff}$. 
For the unimodal cluster in panel 1, $\epsilon_{eff} = \epsilon_{agg}$. For 
the bimodal cluster in panel 2, the estimate of $\epsilon_{eff}$ is weighted 
more by the larger component. It is zero for the case shown in panel 2. This 
is also true for the cluster in panel 3.
However, for a bimodal system with equal sized sub-clumps but different 
elongation, $\epsilon_{eff}$ will give an average elongation of the two. 
For systems with sub-structures the estimate of ${\epsilon}_{agg}$, on the 
other hand, tells us about the overall shape of these systems. Due to the 
presence of two isolated components, the system itself appears more 
elongated than the shape of its sub-clumps. 
An important point to note that the estimate provided by ${\epsilon}_{agg}$ 
depends not only on the relative sizes of the components but also on their 
relative separations. This is reflected in the panels containing multi-clump 
clusters. For equal separation, a bimodal cluster with components similar 
in shape but unequal in size has lower ${\epsilon}_{agg}$ than that of a 
a bimodal cluster with identical shape and size (see the region 
$1.8 < \log_{10} A_S < 2.2$ in panels 2 and 3 in Fig. \ref{toy_ell}). 
In general, as the density and brightness level decreases the clumps get 
bigger and appear closer to one another and ${\epsilon}_{agg}$ gets smaller. 

Note that for a multicomponent system with filamentary structure, 
$\epsilon_{eff} < \epsilon_{agg}$ (panel 4). On the other hand if components 
are distributed around the cluster center, $\epsilon_{eff} > \epsilon_{agg}$ 
(panel 6). The cluster in panel 5 has the unique property that is shown 
separately by clusters in panels 4 and 6. In transition at a lower brightness 
level, the cluster changes its filamentary shape to an extended structure 
where the components are distributed over a region around the center. 
The $\epsilon_{agg}$ profile in panel 8 shows that in the range, 
$2.2 < \log_{10} A_S < 2.8$, the cluster develops two, almost equal size 
clumps that are very close to each other. The cluster in panel 7 follows the 
behavior of a bimodal cluster except that there is jump in between 
$2.6 < \log_{10} A_S < 2.8$ where the cluster changes its structure having 
two unequal size clumps to two equal size clumps. The shape of the cluster in 
panel 9 changes consistently following the merger of its clumps at different 
brightness levels.

In Fig.s \ref{toy_mul} and \ref{toy_ell} we also show the variants of the 
parameters used later in this study. We use circle and star to represent 
the structural parameters, $\bar{M}_{eff}$ and $M_{max}$. The respective 
symbols are also used for the shape parameters $\bar{\epsilon}_{eff}$ 
and $\bar{\epsilon}_{agg}$. Comparing this set of parameters with more 
general $M$ and $\epsilon$ one can easily see how these measures response 
to the alignment of sub-structures and their spatial locations (filamentary, 
extended etc.).
\subsection{Example of Simulated Clusters} 
\begin{figure}
\epsscale{1.10}
\plotone{\figdir/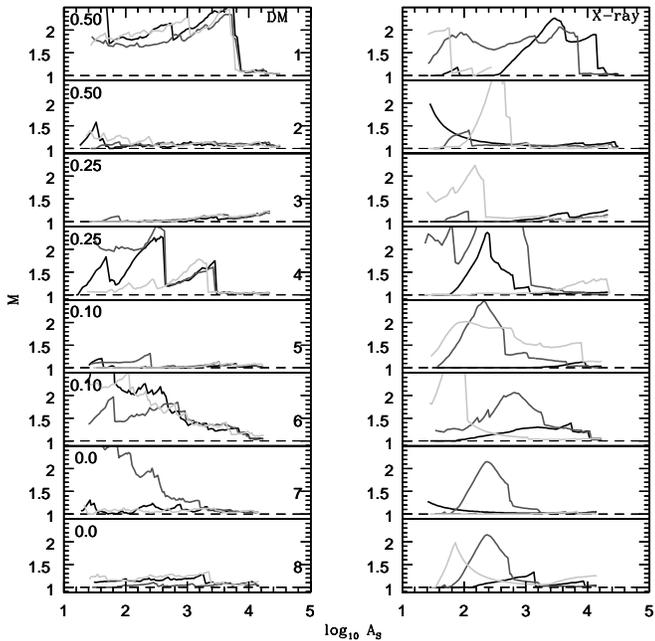}
\caption{Multiplicity (${M}$) as a function of contour area ($A_S$) for 
a selection of clusters at $z = 0.50, 0.25,$ $0.10,$ and $0.0$. Two 
clusters from each redshift are shown. Dark, gray, and faint solid 
lines represent respectively, the adiabatic, radiative cooling (RC), 
and star formation with feedback (SFF) samples. 
The dark matter (DM) and X-ray clusters are shown on the left and right 
panels, respectively. Multiplicity is, in general, greater than 1 in the 
entire redshift range for clusters simulated with RC (medium line) 
indicating a slower evolution than in the adiabatic sample (dark line). 
Redshift $z = 0.5$ is taken for demonstration purpose only. See text for 
details. \label{multip_area}}
\end{figure}

\begin{figure}
\epsscale{1.10}
\plotone{\figdir/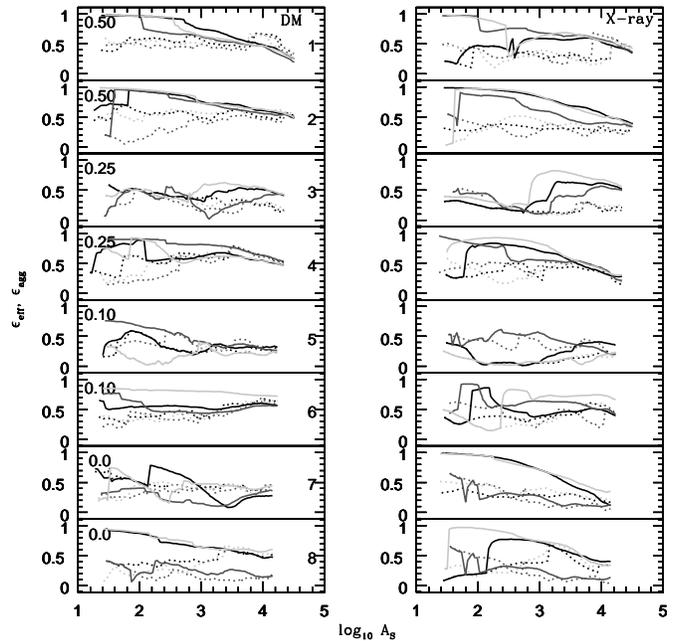}
\caption{Effective (${\epsilon}_{eff}$) and aggregate (${\epsilon}_{agg}$) 
ellipticity as a function of contour area ($A_S$) for the same clusters as 
in Fig. \ref{multip_area}. Solid and dotted lines are used to represent 
${\epsilon}_{agg}$ and ${\epsilon}_{eff}$, respectively. The color style 
is similar to Fig. \ref{multip_area}. In most cases the non-spherical 
central part of these clusters consists of a single peak (i.e. 
${\epsilon}_{eff} = {\epsilon}_{agg}$) whereas in the outer regions 
sub-clumps show various shapes. It can be see easily that the central 
regions of clusters in hydrodynamic simulations appear to be more regular. 
We notice that cluster centers are slighly more flattened than the outer 
parts, irrespective of the nature of simulation. Although the trend is weak. 
See text for details. 
\label{ellip_area}}
\end{figure}

We demonstrate the behaviors of $M$ and the variants of $\epsilon$ as a 
function of area for a collection of simulated clusters in Figs. 
\ref{multip_area} and \ref{ellip_area}. For each sample we choose two 
clusters at each redshift. We use dark, gray, and faint solid lines to 
represent respectively, the adiabatic, RC, and SFF samples. The DM halos 
and X-ray clusters are shown on the left and right panels, respectively. 

Figure \ref{multip_area} shows that both matter and X-ray clusters with 
cooling, generally, have a higher number of sub-clumps than those without 
cooling. Figure \ref{ellip_area} shows that in most cases the central part 
of cluster consists of a single peak (${\epsilon}_{eff} = {\epsilon}_{agg}$). 
The central region of these clusters do not appear spherical, rather this 
region has some degree of flattening. We see that multi-peak systems, 
mostly bimodal clusters with un-equal size sub-clumps 
(${\epsilon}_{eff} < {\epsilon}_{agg}$), are common for these clusters. 
At low brightness levels, i.e. in the outer regions of clusters, the 
sub-clumps appear in various shapes. In some cases they merge forming 
one system (${\epsilon}_{eff} = {\epsilon}_{agg}$), in few cases they 
appear homogeneously distributed (${\epsilon}_{eff} > {\epsilon}_{agg}$), 
and in few cases they form filamentary structure 
(${\epsilon}_{eff} < {\epsilon}_{agg}$). 
The degree of inhomogeneity (${\epsilon}_{eff} \neq {\epsilon}_{agg}$), 
generally, is higher for X-ray clusters. There is a weak trend that cluster 
centers are more flattened than the outer parts, irrespective of the nature 
of simulation. 
\section{Results}
One of the objectives of this paper is to study morphological evolution in 
simulated and optical clusters using $M$ (equation \ref{mul}) and $\epsilon$ 
(equations \ref{ell1} - \ref{ell3}) as quantitative measures. These parameters 
represent the shape characteristics of a set of iso-density/intensity contours 
corresponding to a set of density/brightness levels. The levels represent equal 
interval in area, i.e. size in log space, which allow higher resolution and 
hence higher weight to the dense, central region. 

In this study, we emphasize the morphological properties of the dense, central 
region of clusters. We analyze each cluster at two different threshold levels 
corresponding to radii $\sim 0.5$ h$^{-1}$ Mpc and $\sim 1$ h$^{-1}$ Mpc where 
the outer radius is within $\sim3$ times the core radius (Bahcall 1999). For 
each radius, measurements are relative to the center.

Cluster images are smoothed by a Gaussian filter with a smoothing scale (ss) 
$\sim 50$ h$^{-1}$ kpc. We choose this scale after trials with different 
values. Our experience shows that for a scale smaller than $\sim 50$ h$^{-1}$ 
kpc, images contain too much noise whereas for a larger scale they become 
over smoothed. We note that smoothing affects the gross morphology without 
much distortion in the evolutionary trend of the parameters. The trend holds 
for both simulated and observed cluster samples. 

\begin{figure}
\epsscale{1.10}
\plotone{\figdir/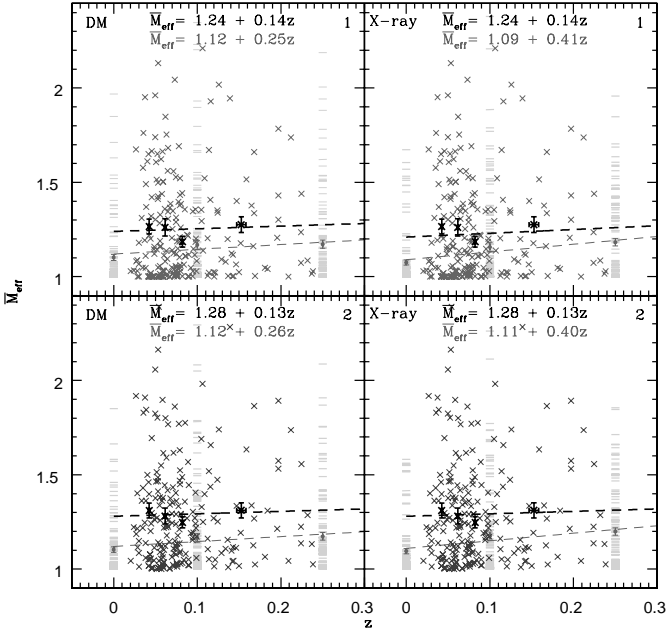}
\caption{A detailed comparison of the estimate of $\bar{M}_{eff}$ for the 
adaibatic DM (left panels) and X-ray (right panels) clusters and the 
optical sample with ss$ = $50 h$^{-1}$ kpc within 0.5 h$^{-1}$ Mpc (panels 
1) and 1.0 h$^{-1}$ Mpc (panels 2) radius. Simulated clusters are shown by 
(faint) horizontal lines at $z = 0.25, 0.10, 0.0$ and the optical clusters 
are shown by (dark) crosses. The expressions represent the best fit lines 
for observation (dark; top line in each panel) and simualtion (faint; second 
line in each panel).
\label{ad_op_mco}}
\end{figure}

\begin{figure}
\epsscale{1.10}
\plotone{\figdir/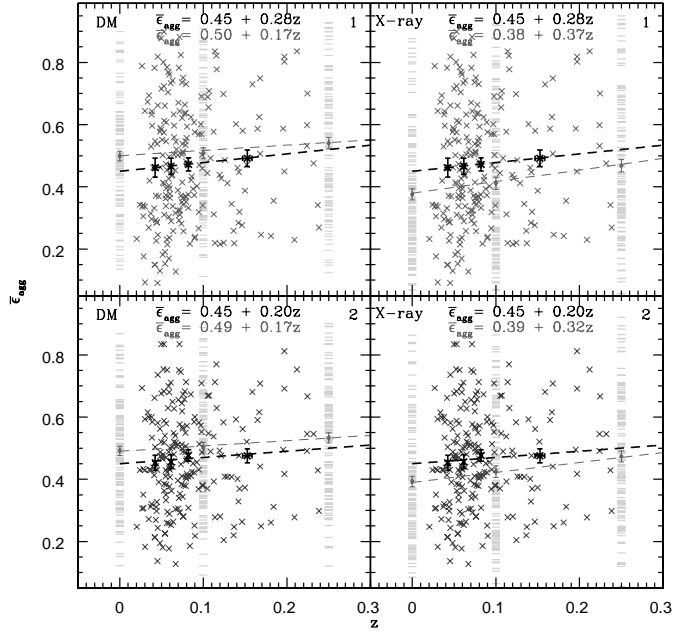}
\caption{A detailed comparison of the estimate of $\bar{\epsilon}_{agg}$ 
obtained from the adaibatic DM (left panels) and X-ray (right panels) 
clusters and from the optical sample. Presentation style is similar to 
Fig. \ref{ad_op_mco}. 
\label{ad_op_eco}}
\end{figure}

In Figs. \ref{ad_op_mco} and \ref{ad_op_eco}, we show detail properties of 
adiabatic and observed clusters using $\bar{M}_{eff}$ and $\bar{\epsilon}_{agg}$.
These figures show results for ss $ = $ 50 h$^{-1}$ kpc within a radius of 
0.5 h$^{-1}$ Mpc (panels numbered 1) and 1 h$^{-1}$ Mpc (panels numbered 2). 
Simulated clusters are shown by (faint) horizontal lines at three different 
redshifts and the optical clusters are shown by (dark) crosses. The expressions 
at each panel represent the best fit line relating the mean of the parameter 
to the redshift although we note that at each redshift the distribution 
functions are highly non-Gaussian.
An interesting feature of these figures is that, at least within $z \leq 0.25$, 
optical clusters have similar dispersion in both $\bar{M}_{eff}$ and 
$\bar{\epsilon}_{agg}$. Similar behavior is also noticed for other parameters 
in this redshift range, irrespective of simulation types. The wide spread in 
multiplicity and projected shape of DM halos and X-ray gas is a clear reflection 
of different merging history (Jing \& Suto 2002). Note that the error bar in the 
normalizations (i.e., intercepts) of the best fit lines is less than $10\%$ for 
all parameters (not shown in these figures). 

We quantify the rate of evolution by the slope of the best fit line where the 
rate means either $d M/d z$ or $d \epsilon/d z$. For gross morphology, we refer 
to the normalization of this line.
We present our final results in Figs. \ref{ad_sm1_ap1} to \ref{sff_sm1_ap2} 
within $\sim 0.5$ h$^{-1}$ Mpc and $\sim 1$ h$^{-1}$ Mpc radii with ss 
$\sim 50$ h$^{-1}$ kpc both for simulated and optical samples. For simulations 
we show the best fit line along with its expression in gray color.
We divide the optical sample into four bins with equal number of clusters in 
each bin. In this case the best fit line is shown in dark color. No expression 
is given for this line. In both simulated and observed clusters the error bar 
represents the error in the mean.
\subsection{Comparison among Simulated Cluster Samples}\label{sec-sim} 
\begin{figure}
\epsscale{1.10}
\plotone{\figdir/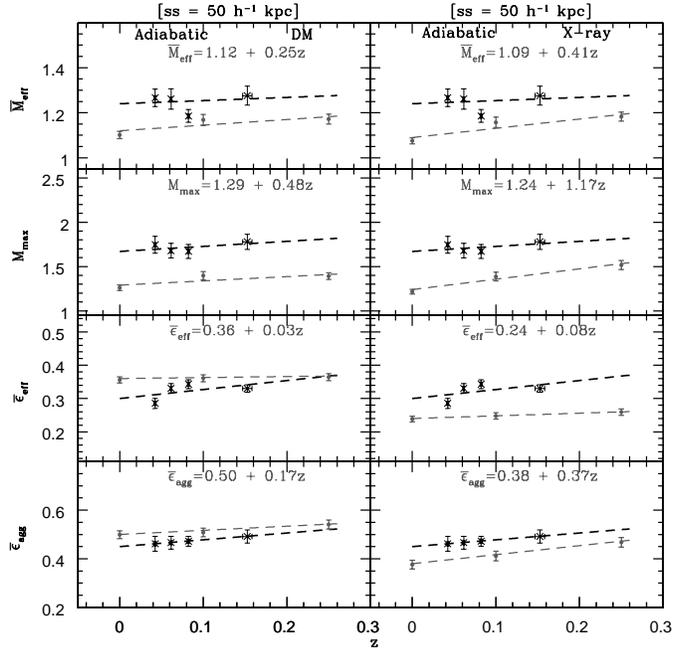}
\caption{Adiabatic sample with 50 h$^{-1}$ kpc smoothing (ss) within 
0.5 h$^{-1}$ Mpc radius. Dark and gray lines are used for optical and 
simulated clusters, respectively. 
The error bar represents the error in the mean. The expression at each 
panel relate the evolution of the mean value of the parameter of with 
redshift. The strength of evolution for optical clusters are: 
$d\bar{M}_{eff}/dz \sim 0.14, \ dM_{max}/dz \sim 0.57,$ 
$d\bar{\epsilon}_{eff}/dz \sim 0.27,$ and $d\bar{\epsilon}_{agg}/dz \sim 0.28$.
\label{ad_sm1_ap1}}
\end{figure}

\begin{figure}
\epsscale{1.10}
\plotone{\figdir/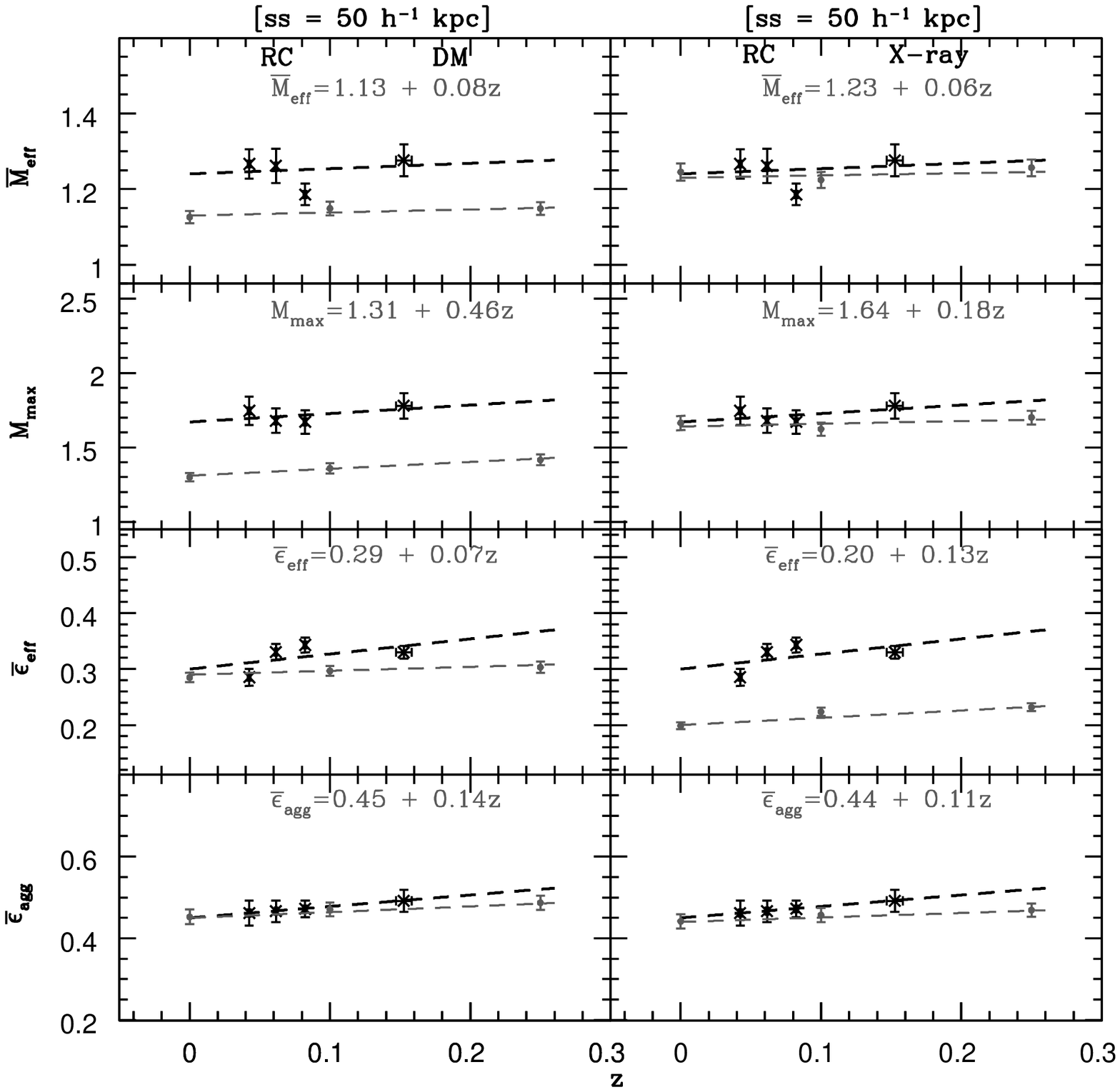}
\caption{Radiative cooling (RC) sample with 50 h$^{-1}$ kpc smoothing 
(ss) within 0.5 h$^{-1}$ Mpc radius. Presentation style is similar to 
Fig. \ref{ad_sm1_ap1}. 
\label{rc_sm1_ap1}}
\end{figure}

\begin{figure}
\epsscale{1.10}
\plotone{\figdir/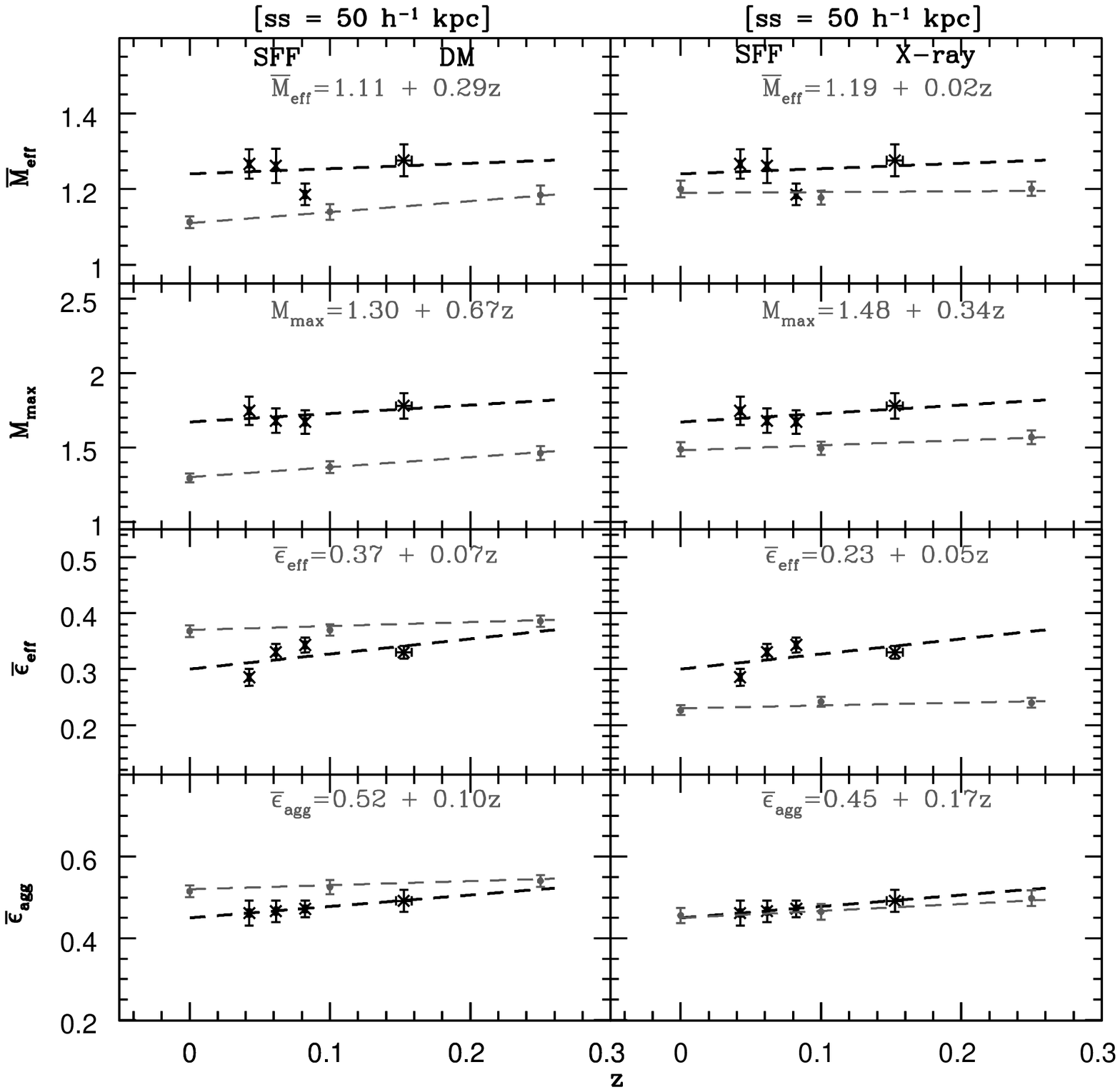}
\caption{Star formation with feedback (SFF) sample with 50 h$^{-1}$ kpc 
smoothing (ss) at 0.5 h$^{-1}$ Mpc radius. Presentation style is similar 
to Fig. \ref{ad_sm1_ap1}. 
\label{sff_sm1_ap1}}
\end{figure}

\begin{figure}
\epsscale{1.10}
\plotone{\figdir/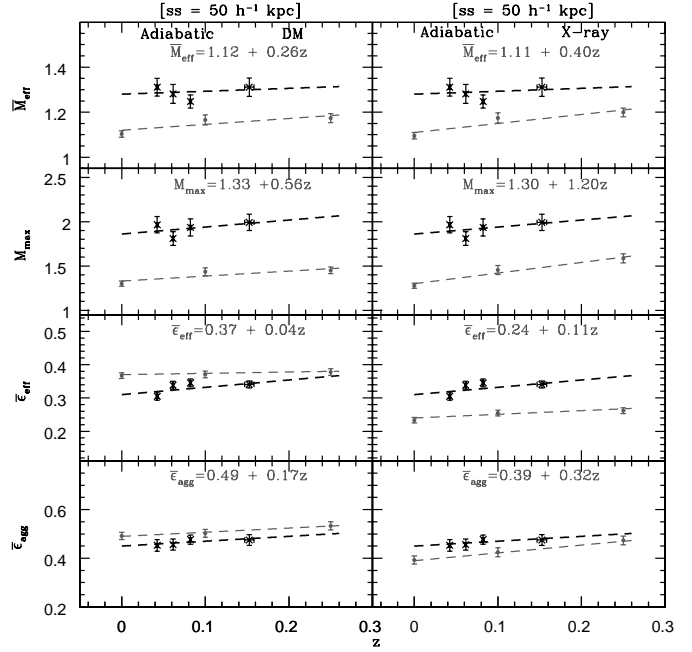}
\caption{Adiabatic sample with 50 h$^{-1}$ kpc smoothing (ss) within 
1.0 h$^{-1}$ Mpc radius. Presentation style is similar to 
Fig. \ref{ad_sm1_ap1}.  The strength of evolution for optical clusters are: 
$d\bar{M}_{eff}/dz \sim 0.13, \ dM_{max}/dz \sim 0.79,$ 
$d\bar{\epsilon}_{eff}/dz \sim 0.22,$ and $d\bar{\epsilon}_{agg}/dz \sim 0.20$.
\label{ad_sm1_ap2}}
\end{figure}

\begin{figure}
\epsscale{1.10}
\plotone{\figdir/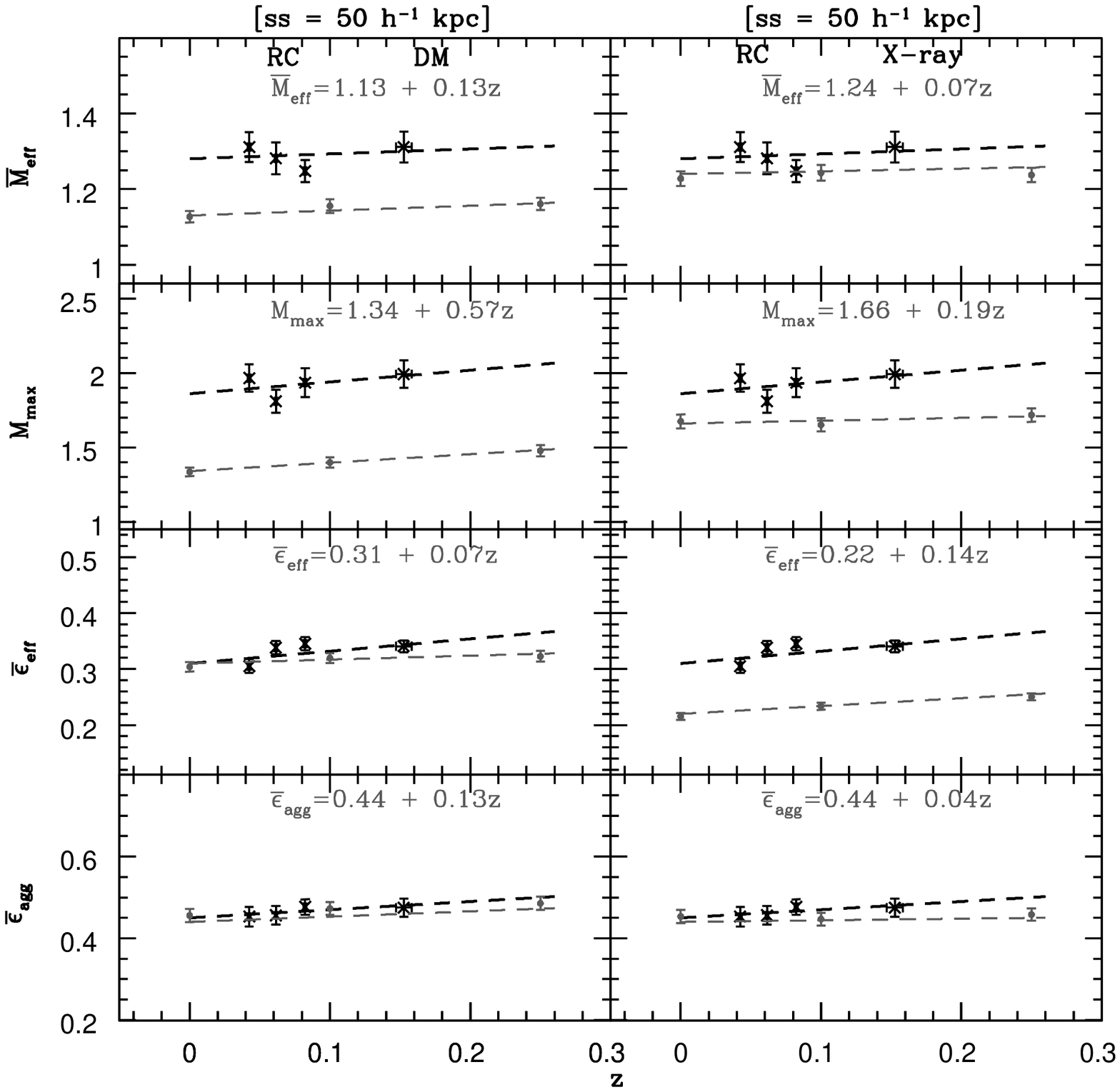}
\caption{Radiative cooling (RC) sample with 50 h$^{-1}$ kpc smoothing 
(ss) within 1.0 h$^{-1}$ Mpc radius. Presentation style is similar to 
Fig. \ref{ad_sm1_ap1}.  \label{rc_sm1_ap2}}
\end{figure}

\begin{figure}
\epsscale{1.10}
\plotone{\figdir/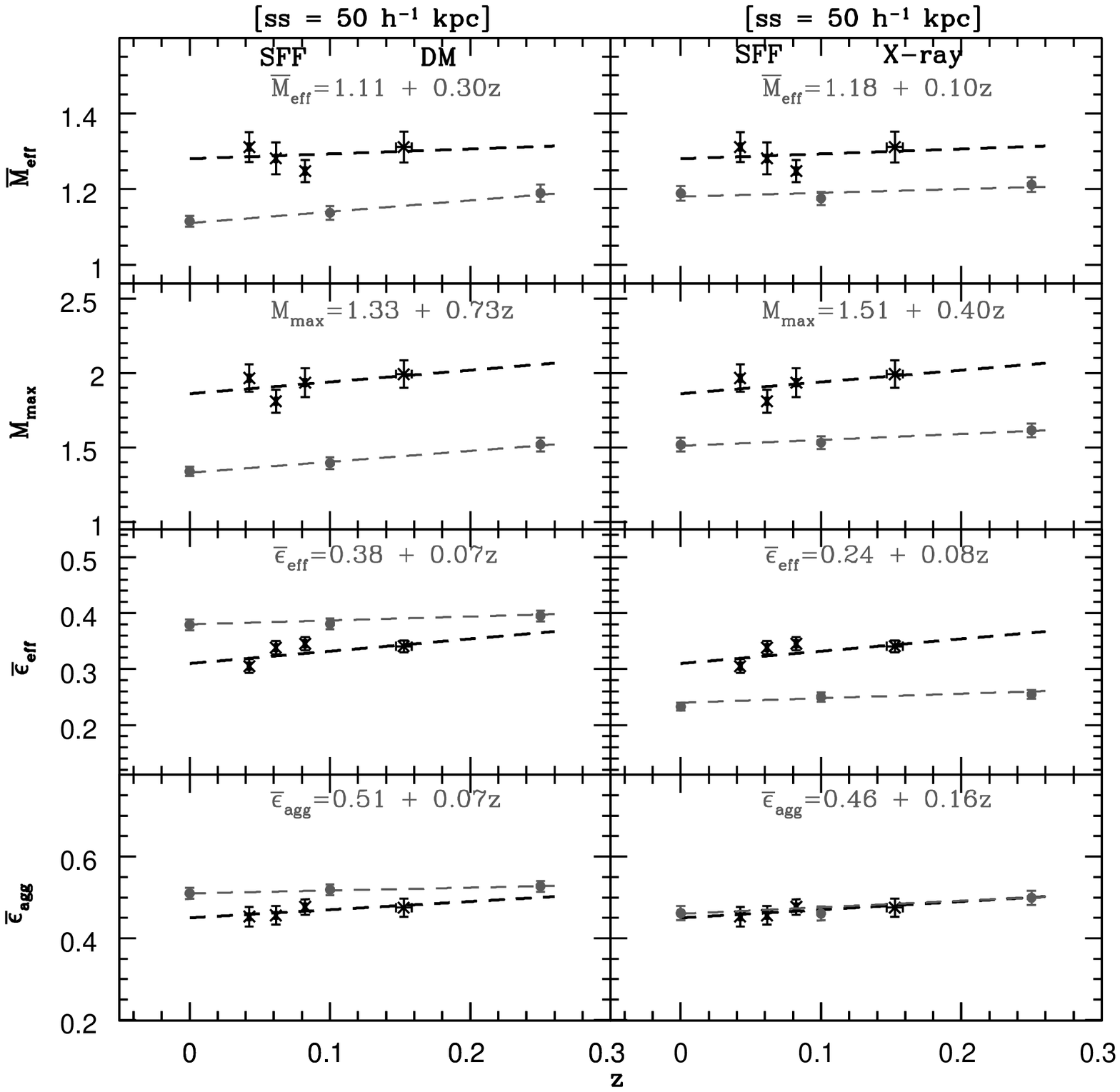}
\caption{Star formation with feedback (SFF) sample with 50 h$^{-1}$ 
kpc smoothing (ss) within 1.0 h$^{-1}$ Mpc radius. Presentation style is 
similar to Fig. \ref{ad_sm1_ap1}.  \label{sff_sm1_ap2}}
\end{figure}

A visual examination of Figs. \ref{ad_sm1_ap1} - \ref{sff_sm1_ap2} clearly 
shows an evolutionary trend in cluster morphology since the gross properties 
of clusters indeed change with redshifts. We are interested to determine the 
significance of this trend of cluster properties computed at two different 
different regions surrounding the cluster center and than to compare it with 
observations. 

Cluster properties within $\sim0.5$ h$^{-1}$ Mpc of the different samples are 
shown in Figs. \ref{ad_sm1_ap1} - \ref{sff_sm1_ap1}. 
In terms of multiplicity, a parameter that probes the number of sub-components 
present in a complex system, we find that cooling samples have slightly higher 
value of multiplicity at all redshifts compared to that in the adiabatic sample. 
The low abundance of single component systems with radiative cooling indicates 
that the dense, cool core sub-structures are long lived features (Motl et al. 
2004). We find that the feedback mechanism with cooling makes cosmological 
systems less clumpy than systems without feedback. 
Energy feedback process, most likely, slows down the rate of evolution in the 
X-ray clusters than those in the cooling only samples. However, this is quite 
opposite for the DM halos. Higher multiplicity in the X-ray clusters in the 
cooling samples indicates a possibility of less efficient merging in hot 
baryonic gas. 

In all simulations multiplicity shows a clear trend with redshift: clusters 
have higher (mean) multiplicity at higher redshifts. Cluster multiplicity 
reflects sub-structures merger rate. It decays by the rate at which the 
cluster can relax, a time scale which is roughly equal to the dynamical time. 
The CDM halos host a larger amount of substructure at higher redshifts because 
of lower accretion time as compared to the dynamical time (see Zentner et al. 
2005). This is the reason for the systematic increase in overall sub-structures 
with increasing redshifts.

In terms of ellipticity, we find that the X-ray clusters, in general, are more 
regular than the halos. The X-ray emitting hot gas is supported by the thermal 
pressure. Due to its isotropic pressure support the X-ray gas becomes 
homogeneously distributed in the background DM potential where it evolves. As 
a result, morphology of the distribution of X-ray gas appears more regular. Our 
results suggest that in X-ray clusters the irregular sub-components are 
distributed over a region instead of making a filamentary structure along one 
direction. A comparison of $\bar{\epsilon}_{eff}$ with $\bar{\epsilon}_{agg}$ 
for the  halos (in all samples) shows that the halo sub-clumps are not 
distributed uniformly around the central region. Rather these clumps are 
spread out mostly in one direction forming filamentary structure, as 
indicated by the larger value of $\bar{\epsilon}_{agg}$. 

No significant evolution is signaled by $\bar{\epsilon}_{eff}$ for the DM 
clusters in any of these samples. Recall that this parameter is an indicator 
of shapes of individual components in a cluster. Therefore, no evolution means 
that shapes of isolated components in clusters at one redshift appear similar 
any other redshifts. Since it places emphasis on individual component, 
therefore, it is not unusual to find no evolution quantified by this parameter. 
However, shapes of sub-clumps in the distributions of X-ray gas change in 
the cooling simulations compared to other simulations.

Properties of simulated clusters within $\sim 1$ h$^{-1}$ Mpc are shown in 
Figs. \ref{ad_sm1_ap2} - \ref{sff_sm1_ap2}. We notice that in this distance, 
small scale structures of simulated clusters do not change significantly than 
what we find in smaller scale. This implies that the small sub-clumps can 
exists up to a Mpc scale and distributed widely over the cluster body. 
However, in this scale sub-structures evolves a bit faster. We find that 
individual, isolated components become a bit more flattened and their 
evolution is slightly stronger. The overall shape of the clusters, however, 
is less flattened than the central region and evolution is weak in all 
simulations.

We take projections along each axis at a time and repeat our analysis. Recall 
that in this case, each sub-sample (along each axis) has only 41 clusters. The 
analysis of these sub-samples do not show significant variation from the primary 
sample. Therefore, it is unlikely that the overall result may have contaminated 
by the projection effect. We have also repeated our analysis using different 
values for the density/brightness levels. Apart from a minor change in gross 
morphology, we find similar results for the rate of evolution. 

We summarize our main results as follows: First, the DM halos show very similar 
evolution in all samples of clusters. Second, the X-ray clusters in the 
adiabatic simulation evolve faster than those with radiative cooling. 
Third, morphology of the central parts of clusters evolve slightly strongly 
than the outer regions. Fourth, feedback processes with cooling makes the DM 
halos slightly more flattened and slower in evolution than the cooling only 
simulations (see Kazantzidis et al. 2004 for a similar trend). 

We emphasize that the measured quantities for the DM distributions in all three 
samples are very similar. This is a check on the consistency of the simulations 
and analysis. The result is expected as the N-body segment of the simulations 
are identical in all three cluster samples with the exception of the gas that 
makes a relatively minor contribution to the total gravitational potential. 
The LSS of adiabatic and cooling clusters are generally similar but their small 
scale structures are determined by the overall cluster properties rather than 
perturbative interactions (Motl et al. 2004). In the adiabatic clusters, the 
mixing of in-falling sub-clumps into the main cluster medium is quicker relative 
to the radiative cooling clusters where sub-structures can be long lived. This 
is a reason behind the fast evolution of adiabatic X-ray clusters. The relaxation 
time scale for collisionless particles is much longer than that of the collisional 
gas particles (Frenk et al. 1999; Valdarnini, Ghizzardi \& Bomometto 1999). 
Therefore, the DM halos will appear not only more elongated than the distributions 
of X-ray gas, but the redshift evolution of their shapes will also be slower. 
More spherical configurations for X-ray clusters is also expected from the point 
of view that intra cluster gas is approximately in hydrostatic equilibrium and 
supported by isotropic pressure (Sarazin 1988). The DM, on the other hand, 
appears to be distributed like galaxies as indicated by recent observations 
from gravitational lensing (Fischer et al. 1997; Fischer \& Tayson 1997; 
Kochanek 2001; Hoekstra 2003; Hoekstra et al. 2004) and by high resolution 
hydrodynamical simulations (Nagai \& Kravtsov 2004; Kang et al. 2005; Maccio 
et al. 2005). 

There is a consensus based on the observations of X-ray clusters that cooling 
affects the mass distribution appreciably only in the inner $\sim 10\%$ of 
the virial radius of cluster size halos (Sarazin 1986). Contrary to that recent 
high resolution hydrodynamic simulations show something quite interesting 
(see Kazantidis et al. 2004). Kazantidis et al. 
show that there is a significant difference in overall shape between 
dissipationless and dissipative simulations which is persistent up to the 
virial radius. The virial mass of their DM halos ranges from 
$\approx 10^{13}$ to $3 \times 10^{14}$ h$^{-1}$ $M_{\odot}$ which translates 
to the virial radius range $\sim 0.26 - 0.82$ h$^{-1}$ Mpc assuming 
$\Delta_{vir}(z=0) \sim 337$, $h \sim 0.7,$ and $\rho_{c} \sim 1.87 h^2 \times 
10^{-29} $ gm cm$^{-3}$ (Kolb \& Turner 1990; Zentner et al. 2005). Kazantidis 
et al. present their analysis upto the virial radius. However, from the trend 
seen right at the virial radius, it seems likely that it goes a bit further 
down along the radial direction before shapes in dissipative and dissipational 
simulations converge. 

Baryon fraction ($\Omega_b \sim 0.043$) in Kazantidis et al. simulations is 
larger than what has been used in our simulations. Therefore, question can be 
raised whether low baryon density can also produce systematic shift in the 
shapes of DM halos to be robust on scales of Mpc as noted in our work. Our 
simulations use baryon density ($\Omega_b = 0.026$) and normalization of 
fluctuation spectrum ($\sigma_8 = 0.928$) which are slightly off than the 
corresponding WMAP values ($\Omega_b=0.044$ and $\sigma_8 = 0.84$, Spergel et 
al. 2003). Baryon density is an important cosmological parameter which affects 
radiative cooling and X-ray luminosity at the central region of large virialized 
structures. Higher $\Omega_b$ enhances the cooling rate, 
subsequently making the central region more regular (Sarazin 1986, Kazantzidis 
et al. 2004, Springel \& White 2004, Allgood et al. 2005, Flores et al. 2005). 
Recent numerical simulations show that larger $\sigma_8$ produces DM halos that 
are more regular in the central regions (Allgood et al. 2005). Therefore, we 
note that cooling is under-emphasized while the core DM sub-structure is 
over-emphasized is our simulations. It may be likely that the offset of 
$\Omega_b$ and $\sigma_8$ compared to WMAP would balance each other and our 
results obtained from the cooling simulations would still be representative 
had we been using the WMAP values.  

With cooling only, our simulated clusters of galaxies show a large amount of 
long-lived substructure compared to the other simulated samples.  While the 
amount of cooling in this sample is unphysical it represents an interesting, 
theoretical, limiting case. On the scale of the cluster itself, the 
gravitational and dynamical effects of cool, dense cores of gas have 
significantly altered the shape of the clusters to length scales comparable 
to the virial radius (see Figs. \ref{multip_area} \& \ref{ellip_area}). The 
perturbation from cool baryonic clumps may thus significantly alter model 
dependent  mass maps derived from weak lensing studies. The robust substructures 
present in the cooling only sample may also play a role in steepening the total 
cluster mass profile (Maccio et al. 2005) and with higher resolution simulations 
may bound the possible contribution of substructures to strong lensing in 
clusters. Though beyond the scope of the current paper, these connections to 
lensing studies will be pursued in future work.

Numerical simulations provide interesting information on two different aspects 
of the LSS: 1) shape of the central structures in galaxy or cluster size halos, 
and 2) Change in shape of halos with radial distance, irrespective of the 
nature of simulation. 
Recent high resolution hydrodynamical simulations has shown quite successfully 
that hydrodynamical phenomena make cluster centers considerably more 
spherical than those in the adiabatic simulations. However, radial dependence of 
shape is still a controversial issue. While Frenk et al. (1988), Bullock (2002),  
Springel, White \& Hernquist (2004), Hopkins, Bahcall \& Bode (2005), and hydro 
simulation of Kazantidis et al. (2004) agree that inner part of clusters are 
more spherical than the outer part, the following groups of Dubinski \& 
Carlberg (1991), Warren et al. (1992), Jung \& Suto (2002), Allgood et al. 
(2005), and hydro simulation of Tissera \& Dominguez-Tenreiro (1998) find it 
completely opposite. Our results closely follow the latter group.  

Note that radial dependence of shape is not monotonic. It changes in a quite 
complicated way depending on the presence of sub-clumps as one can see from Fig. 
\ref{ellip_area}. A similar trend is also seen in hydro simulations of 
Kazantzidis et al. (2004).
\subsection{Comparison with Optical Clusters}\label{sec-opt}
We have analyzed a sample of ACO clusters within redshift, $z \leq 0.25$. 
The sample contains 208 optical clusters derived from 10-inch photographic 
plates taken with the 48-inch Palomar Schmidt Telescope (for details of 
the data acquisition and processing see Tr\`{e}vese et al. 1992; Flin et 
al. 1995; Tr\`{e}vese et al. 1997; Flin et al. 2000). Results obtained from 
the  optical clusters are shown in Figs. \ref{ad_sm1_ap1} - \ref{sff_sm1_ap2} 
using dark dashed lines. The summary of our results is as follows:
 
$\bullet$ The optical clusters are, in general, more clumpy than the 
simulated DM halos as given by both $\bar{M}_{eff}$ and $M_{max}$. 
The sub-structure at the central part of X-ray clusters in the RC sample 
are compatible with the optical clusters, at least, within redshift 
$z \leq 0.25$. At large radius, the optical clusters include more small 
scale structures and show stronger evolution in sub-structures. 

$\bullet$ The sub-structures of hot baryonic gas 
evolve much strongly in the adiabatic simulation than that in the galaxy 
distribution ($d \bar{M}_{eff} / d z \sim 0.14, 0.13$ in 0.5 and 1 h$^{-1}$ 
Mpc). On the other hand, effective sub-clumps ($\bar{M}_{eff}$) of the halos 
have faster rate in both adiabatic and SFF simulations. Feedback process 
along with radiative cooling make rapid evolution in DM halo structures. In 
terms of $M_{max}$, however, evolution of the galaxy distribution is always 
stronger compared to all three simulations 
($d M_{max} / d z \sim 0.57, 0.79$ for 0.5 and 1 h$^{-1}$ Mpc, respectively).

$\bullet$ The largest component of the DM halos (probed by $\bar{\epsilon}_{eff}$) 
in the adiabatic and SFF simulations have higher elongation compared to that in 
the galaxy distribution. In the RC simulation we find an opposite trend. The 
shape of the largest sub-clump formed in the distributions of X-ray emitting 
hot gas in all three simulations are significantly rounder than that of the 
optical clusters ($d \bar{\epsilon}_{eff} / d z \sim 0.27, 0.22$ in 0.5 and 1 
h$^{-1}$ Mpc, respectively).  

$\bullet$ The overall shape (probed by $\bar{\epsilon}_{agg}$) and the 
strength of evolution in optical clusters 
($d \bar{\epsilon}_{agg} / d z \sim 0.28, 0.2$ in 0.5 and 1 h$^{-1}$ Mpc, 
respectively) show nice agreement with that of the 
X-ray clusters in dissipative simulations. In dissipationless simulation, 
however, hot gas is systematically less elongated but evolve much strongly
than the galaxy distribution ($d \bar{\epsilon}_{agg} / d z \sim 0.28, 0.2$ 
in 0.5 and 1 h$^{-1}$ Mpc, respectively). 

$\bullet$ The shapes of the optical clusters are comparable to the halos only 
in the RC simulation. The halos are slightly more flattened and slower in 
evolutionary process compared to the galaxy distributions in the adiabatic 
case and in simulation including feedback processes with cooling.

$\bullet$ The strength of shape evolution given by $d\bar{\epsilon}_{agg}/dz$ 
is slightly stronger around the cluster core, in both observed and simulated 
clusters. 

There are several possibilities for optical clusters to be more clumpy. First, 
the choice of smoothing may not be optimal for the optical sample. The smoothing 
scale used in our study, therefore, should be taken as the lower limit. Second, 
projection effect due to the background galaxies may also play an important role. 
This effect becomes significant as one moves away from the clusters center 
(Kolokotronis et al. 2001). 
Third, since ACO clusters are selected via richness criteria and it has been 
shown that richness is poorly correlated with mass (Girardi et al. 1998; Miller 
2004, private communication), there is a chance that optical sample may be 
biased toward high mass end of cluster mass range or clusters that have gone 
through recent merger. Massive clusters are dynamically less relaxed and hence 
rich in sub-structure.
A well defined mass selection criteria needs to be applied for a more systematic 
comparison as recent numerical simulations show that cluster shapes depend on 
mass, although the mass - shape correlation is weak and show large dispersion 
(Bullock 2002; Jing \& Suto 2002; Hopkins, Bahcall \& Bode 2005; Allgood et al. 
2005). 

Regarding the comparison of the strength of evolution we note that the 
morphological parameters derived for the set of simulated halos have uniform 
statistical weight at all redshifts. However, this is certainly not the case 
for the observed sample as it has considerably more weight toward $z = 0.0$ 
than the simulated samples (see Fig.s \ref{ad_op_mco} and \ref{ad_op_eco}). 
As mentioned earlier the scatter is comparable in both samples and, therefore, 
we believe the choice of binning has less effect on the overall outcome of our 
analysis.
\subsection{Results from Previous Studies on Observed Clusters}\label{sec-pre}
In this section we summarize the results obtained from previous studies on 
optical and X-ray clusters. Our objective is to highlight the fact that 
different samples of clusters give different rates of evolution in cluster 
morphology. Due to the methodological differences, we refrain from making 
a direct comparison with the results of these studies.

To find cluster shapes, all previous studies follow the procedure described 
in Carter \& Metcalf (1980). These studies, however, differ in adopting 
weighting factor, threshold level, cluster center and smoothing techniques 
to construct galaxy density distribution from spatial distribution. It is 
also important to note that cluster shape quantified by ellipticity is not 
uniquely defined. Therefore, to help reader getting a better feelings about 
the inherent differences of previous studies we also provide a brief outline 
of the methodologies used in these studies.  

The optical sample of MCM contains 138 ACO clusters with $z < 0.1$ which 
has been compiled from West \& Bothun (1990); Rhee, van Haarlem \& 
Katgert (1991) and Kolokotronis et al. (2001). This sample show no 
significant evolution, $d \epsilon/d z \sim 0.03$. 

The former two groups measure cluster shapes from discrete galaxy 
distribution using method of moments. They define the two-dimensional moments 
as, 
\begin{equation}
\mu_{mn} = \frac{\sum_{i,j} \ (x_i-x_0)^m (y_j-y_0)^n}{N},
\end{equation}
where $x_0, y_0$ are the coordinates of the brightest galaxy taken as the 
cluster center, $N$ is total number of galaxies within the region which is 
$3\sigma$ above the background noise and $m, n =0, 1, 2$. They diagonalize 
the matrix formed by the components $\mu_{20}, \mu_{02},$ and $\mu_{11}$, 
find the eigenvalues and obtain cluster shape using eigenvalues from the 
relation, 
$\epsilon = 1 - \lambda_2^2/\lambda_1^2$, where $\lambda_1 > \lambda_2$. 
Kolokotronis et al. (2001) use moment of inertia method for a sample 
containing 22 APM clusters along with their ROSAT counterparts in the 
redshift range, $z \leq 0.13$. They use Gaussian smoothing on galaxy 
density distribution and define the components of the symmetric inertia 
tensor as,
\begin{eqnarray}
I_{11} = \sum_{i} w_i (r_i^2-x_i^2), \ I_{22} = \sum_{i} w_i (r_i^2-y_i^2), 
\nonumber \\ 
I_{12} = I_{21} = -\sum_{i} w_i x_i y_i,
\end{eqnarray}
where $w_i$ is the average cell density within $0.75$ h$^{-1}$ Mpc region 
and $r_i^2 = x_i^2 + y_i^2$. After defining inertia tensor, Kolokotronis et 
al. follow similar route to the other groups to define shape except that 
they define ellipticity as, $\epsilon = 1 - \lambda_2/\lambda_1$. 

Plionis (2002) analyze the largest sample of optical clusters following the 
method used in Kolokotronis  et al. (2001). His sample has 407 APM clusters 
within a volume of $z < 0.18$. The rate of evolution for the Plionis sample 
is $d \epsilon/d z \sim 0.7$. However, if both are combined, replacing the 
common ones by the APM clusters, the rate increases. The combined sample of 
$\sim 500$ optical clusters with $z < 0.18$ shows $d \epsilon/d z \sim 1.06$. 

It is rare to find a large sample of X-ray clusters with up-to-date 
ellipticity measurements. The X-ray sample of MCM is compiled from Mcmillan, 
Kowalski \& Ulmer (1989; hereafter MKU) and Kolokotronis et al. (2001). MKU 
measure cluster shape using method of moments from 2D X-ray surface brightness 
images. They adopt the following definition of the moment,   
\begin{equation}
\mu_{mn} = \frac{\sum_{i,j} \ f_{ij} \ (x_i-x_0)^m (y_j-y_0)^n}{\sum_{ij} f_{ij}}.
\end{equation}
where $x_0, y_0$ are the components of the image centroid,
$x_0 = \sum x_i f_{ij}/\sum f_{ij}$ and $y_0 = \sum y_j f_{ij}/\sum f_{ij}$. 
They determine the overall shape of a cluster using the faintest flux level 
available for that object. This sample has 48 clusters with $z < 0.1$ which 
is three times smaller than the MCM optical sample and an order of magnitude 
smaller than the APM sample. It also has a lower redshift limit than the APM 
sample. 
The rate of evolution for this sample is $d \epsilon/d z \sim 1.7$. The result 
suggests faster evolution for the X-ray clusters than the optical one. 
Interestingly, a comparison of optical and X-ray clusters within 
Kolokotronis et al. (2001) sample show completely opposite trend: galaxy 
density distributions have stronger evolution than the distribution of hot 
gas. The galaxy and X-ray cluster shapes follow a trend where flattened gas 
distribution signals anisotropic distribution of galaxies. However, the 
scatter is large in both relationships. It is not clear to us what could be 
the reasons of possible contradictions except the fact that MCM sample is 
most likely contaminated due to different methodologies. Besides it is also 
difficult to make any definite conclusion because of the smaller sizes of 
the samples. A large sample of X-ray clusters with better selection criteria 
and extended to higher redshift is needed.
 
We reanalyze $\epsilon-z$ estimates derived from the APM cluster data and 
the combined sample imposing a redshift cutoff $z < 0.1$ in order to be 
consistent with the redshift range of MCM X-ray sample. For these samples, 
we find the rate of evolution as , $d \epsilon / d z \sim 1.02$ and 
$\sim 1.0$, respectively. We find that evolution of optical clusters 
accelerates in this redshift range but it is still slower than that of the 
X-ray.

Flin, Krywult \& Biernacka (2004; hereafter FKB) has analyzed a sample 
of 246 ACO clusters for $z \leq 0.31$. This group use the same definition 
of moment as in equation 1. They use density peak as the cluster center 
and measure shapes at different circular aperture radii ranging from 0.5 
to 1.5 h$^{-1}$ Mpc with an increment of 0.25 h$^{-1}$ Mpc. They estimate 
cluster shape at all radii and find no dependence of cluster ellipticity 
on redshift. Interestingly FKB noted a decrement of $d {\epsilon} / d z$ 
with radius. They find positive evolution at radii of 0.5 and 0.75 h$^{-1}$ 
Mpc. However, for radii $\geq$ 1 h$^{-1}$ Mpc, they report negative 
evolution. The mean of their estimates derived from these five radii shows 
$\bar{\epsilon}$ $\approx 0.22$ and $d \bar{\epsilon} / d z \sim 0.013$. 
For $z < 0.1$, their result also indicates weak evolution.
We use this sample of optical clusters for our analysis (see $\S$4.2) but 
with a reduced number (208) of clusters. The reduction is made after visual 
inspection and it is due to the removal of clusters images that appear either 
small or close to the boundary.

It should be noted that MCM and APM samples emphasize cluster morphology in 
two different regions. The MCM sample excludes any study with radius less 
than 1 h$^{-1}$ Mpc and includes the estimate of ellipticity within 
$\sim$1 - 2 h$^{-1}$ Mpc from the cluster center. The APM sample, however, 
provide information on cluster shape within 0.75 h$^{-1}$ Mpc of the center. 
Therefore, care must be exercised in interpreting and comparing results of 
observed clusters with simulations if both are not analyzed under the same 
measurement technique. Unfortunately the studies of Floor et al. (2003) and 
FMM (2004) has ignored this fact.

In spite of differences in the evolution of cluster morphology, optical 
samples are consistent with one another atleast in one case: shape of galaxy 
density distributions evolve strongly in the central region (Plionis, 2002) 
than that in the outer part (MCM, 2001). Interestingly our results are also 
consistent with this trend. For X-ray clusters this trend has yet to establish. 
\section{Conclusions}
Numerical simulations provide an unique opportunity to follow the hierarchical 
nature of the LSS formation in both linear and nonlinear regimes (Frenk et al. 
1985, 1988; Quinn, Salmon \& Zurek 1986; Efstathiou et al. 1988). In order to 
be representative of the reality, results from simulations should agree 
with observations. Observations provide evidence of morphological evolution in 
galaxy-clusters (Melott, Chambers \& Miller 2001; Plionis 2002; Jeltema et al. 
2005), simulations should show similar trend. 
Besides, in the CDM model luminous galaxies are associated with the DM sub-halos 
which reside in bigger parent halos, closely associated with galaxy clusters. 
According to this model statistical properties of galaxies, e.g. mass, 
sub-structure, shape etc., would show a similar trend to that of the sub-halos 
while X-ray emitting hot gas would have different properties than galaxies and 
sub-halos. A statistical analysis of various properties of halos, galaxy clusters, 
and X-ray gas could provide clues to find possible biasing of luminous galaxies 
toward DM sub-halos and whether or not they have any correspondence with the 
distribution of hot gas. With this in mind, we have 
studied redshift evolution of cluster morphology simulated, respectively, in 
the adiabatic limit, with radiative cooling, and with star formation including 
SN feedback at three different redshifts, $z = 0.0, 0.10,$ and $0.25$. 
For comparision we have also studied a sample of observed clusters containing 
208 ACO clusters within redshift, $z \leq 0.25$.

Since observed clusters are projected along the line of sight and lack the full 
three dimensional information we, therefore, use projected simulated clusters. 
Each cluster image is a 8 h$^{-1}$ Mpc frame containing 360 $\times$ 360 pixels. 
Clusters are analyzed at two different density/brightness threshold levels 
corresponding to radii 0.5 and 1 h$^{-1}$ Mpc from the cluster center. To 
quantify morphological evolution we use multiplicity and ellipticity as two 
different probes that are sensitive to cluster sub-structures and shape. 

Our results indicate that optical clusters have, in general, more sub-structures 
than simulated halos and X-ray brightness distributions. Cluster components, in 
both observed and simulated clusters, evolve with redshifts and the evolution is 
different at different regions from cluster centers.
In terms of total multiplicity ($M_{max}$), observed clusters have stronger 
evolution compared to DM halos. The X-ray brightness distributions, however, 
show steeper evolution (than that of galaxy clusters) in dissipationless 
simulation.

We find that in terms of overall shape, simulations do model the observed 
universe in an interesting way. The simulated clusters evolve with redshift, 
consistent with the hierarchical formation scenario. However, observed 
clusters appear to be slightly more flattened at higher redshift than the 
simulated one indicating slower evolution in simulated objects. This may 
reflect some form of incompleteness in our understanding in simulating the 
LSS. 
Our results differ from those of FMM (2004) who reported that the evolution 
in the simulated cluster shape is significantly slower than the observed one. 
We not only find stronger structural evolution in simulated clusters, but also 
find that observed cluster shapes appear to be consistent with dissipative 
simulations, at least, in the redshift range $z < 0.1$. The discrepancies 
noted in FMM is due to the different redshift ranged probed as well as 
intrinsic methodological differences while comparing simulations with 
observations.  

We note that on one hand shapes of optical clusters seems to be compatible 
with both the halos and X-ray brightness distributions, one the 
other hand, both of these components appear to be less clumpy than the 
distribution of galaxies. Therefore, it seems puzzling whether or not there 
is any correspondence between the DM halos and galaxies.  
The existence of any such correspondence is still a matter of ongoing debate 
as there are conflicting results based on systematics of numerical simulations 
such as nature of simulations (dissipationless or dissipative) and the effect 
of mass and force resolution (see Maccio et al. 2005). In the context of the 
CDM model we would expect that the optical clusters would have similar 
morphology and evolutionary trend to that of the halos and would be different 
than the properties of the distribution of hot gas traced in the X-ray region 
of the spectrum.

Within the uncertainties and systematics involved in our optical sample, the 
results indicate that the properties of optical clusters do not exactly 
represent either the distribution of the halos or that of the X-ray emitting 
gas in any of the simulations. We find offsets in the measured parameters, 
such as multiplicities and ellipticities, between observations and simulations, 
and are unable to find any clear signature of DM-galaxy biasing based on our 
morphological analysis. This may be an indication, although in no way 
conclusive, of the fact that these components of the LSS may represent 
intrinsically different populations, and galaxies may not trace the DM 
distributions (see Gao et al. 2004a,b; Nagai \& Kravtsov 2005). However, 
this is merely a speculation and we stress that care must be exercised in 
interpreting our results as one must be careful in selecting proper measures, 
radius, mass range, and most importantly, well defined samples of clusters 
to have unbiased and meaningful results in any morphological analysis 
comparing observations and simulations.

We find that the measurements from different samples do not agree on the 
evolution rate. Take, for example, optical clusters with $z < 0.1$, and radius, 
0.75 h$^{-1}$ Mpc. In this case, the APM sample shows $d\epsilon/dz \sim 1.02$. 
FKB, on the other hand, finds much weaker evolution, $d \epsilon / d z \sim 0.2$. 
As mentioned in FKM, the discrepancy may be due to differences in adopting 
cluster centers, smoothing, and applied method of shape determination.

A preliminary analysis of a sample of 800 clusters constructed from the Sloan 
Digital Sky Survey (SDSS) shows that ellipticity evolution of optical clusters, 
for $z < 0.1$ and within $\sim 1$ h$^{-1}$ Mpc, is weaker than that of the APM 
clusters. The result indicates that clusters with different mass limits evolves 
differently. Large, massive clusters ($M \sim 10 ^{15} M_{\odot}$) have stronger 
evolution compared to the less massive clusters ($M \sim 10^{13}-10^{14} M_{\odot}$) 
(C. Miller 2004, private communication). This is an interesting observation. If 
it is confirmed then the scaling relation between axes ratios and mass noted in 
simulations (Bullock 2002; Jing \& Suto 2002) must be modified to be consistent 
with observations. 

The SDSS sample is uniform with a well documented selection function and high 
degree of completeness. We may then infer that the cluster samples discussed 
previously have less uniformity in mass range: the APM catalog and FKB samples 
are biased toward massive clusters whereas the MCM samples contain more less 
massive clusters. The discrepancy may also arise from the techniques applied 
in ellipticity estimates (see also Flores et al. 2005 in this regard). 
Unfortunately we are unable to check the evolution strength - mass relation for 
our optical sample because, apart from an approximate range, no well defined 
criteria has been used to sort clusters into different mass bins.

The discrepancy in the optical samples is an indication of different selection 
criteria used to construct the catalogs. Larger and more complete catalogs 
obtained from the SDSS and XMM-Newton survey may be able to shed more light into 
this issue. It is also likely that numerical simulations may lack crucial physics 
that needs to be included (see FMM for discussion). In the future we will analyze 
clusters simulated with various gas physics, e.g. thermal conduction and AGN 
heating, and compare them with the SDSS clusters. The results of these studies 
may give us some clues to gain better insight of the current discrepancy.
\\
\\
\noindent{\bf Acknowledgments}
We thank the anonymous referee for constructive comments and criticisms which 
help to improve the quality of this paper. We thank M. Plionis for providing 
the APM cluster ellipticity data and Scott W. Chambers for the MCM data sets. 
NR thanks Hume Feldman and Bruce Twarog for many useful discussions. 


\begin{thebibliography}{}
\bibitem[\protect\citename{Abell, Crowin \& Olowin 1989}]
{abell-etal89} 
Abell G. O., Crowin H. G., Olowin H. G., 1989, ApJS, 70, 1 (ACO)

\bibitem[\protect\citename{Allgood et al. 2005}]
{allgood-etal05} 
Allgood et al., 2005, MNRAS, submitted, astro-ph/0508497

\bibitem[\protect\citename{Aninos \& Norman 1996}]{ani-nor96}
Aninos P., Norman M. L., 1996, ApJ, 459, 12

\bibitem[\protect\citename{Bahcall 1999}]{bahcall99}
Bahcall N. A., 1999, in Formation of Structure in the Universe, 
eds. A. Dekel \& J. P. Ostriker (Cambridge University Press, New 
York), p. 135

\bibitem[\protect\citename{Beisbart 2000}]{beisbart00}
Beisbart C., 2000, Ph.D. Thesis, 
Ludwig Maximilians Universit\"{a}t, M\"{u}nichen, Germany

\bibitem[\protect\citename{Beisbart, Buchert \& Wagner 2001}]
{beisbart-etal01a}
Beisbart C., Buchert T., Wagner H., 2001, Physica A, 293, 592B

\bibitem[\protect\citename{Beisbart, Valdarnini \& Buchert 2001}]
{beisbart-etal01b}
Beisbart C., Valdarnini R., Buchert T., 2001, A\&A, 379, 412

\bibitem[\protect\citename{Bryan, Abel \& Norman 1999}]
{bryan-etal99}
Bryan G. L., Abel T., Norman M. L., 2001, Proceedings of 
Supercomputing, http://www.sc2001.org/

\bibitem[\protect\citename{Bullock 2002}]{bullock-02}
Bullock J. S., 2002, in The Shapes of Galaxies and Their Dark 
Matter Halos, ed. Priyamvada Natrajan (World Scientific, Sigapore), 
p. 109, astro-ph/0106380

\bibitem[\protect\citename{Buote \& Tsai 1995}]{bou-tsa95}
Buote D. A., Tsai J. C., 1995, ApJ, 452, 522

\bibitem[\protect\citename{Buote \& Xu 1996}]{bou-xu95}
Buote D. A., Xu G., 1996, MNRAS, 284, 439

\bibitem[\protect\citename{Buote et al. 2002}]{boute-etal02}
Buote D. A., Jeltema T. E., Canizares C. R., Garmire G. P., 
2002, ApJ, 577, 183

\bibitem[\protect\citename{Carter \& Metcalfe} 1980]{car-met80}
Carter D., Metcalfe N., 1980, MNRAS, 191, 325

\bibitem[\protect\citename{Cen \& Ostriker 1992}]{cen-ost92}
Cen R. \& Ostriker J. P., 1992, ApJ, 393, 32

\bibitem[\protect\citename{Chandran \& Cowley 1998}]{cha-cow98}
Chandran B. D. G., Cowley S. C., 1998, Phys. Rev. Lett. 80, 3077

\bibitem[\protect\citename{Colella \& Woodward 1984}]{col-woo84}
Colella P., Woodward P. R., 1984, J. Comput. Phys., 54, 174

\bibitem[\protect\citename{Crone, Evrard \& Richstone 1996}]
{crone-etal96}
Crone M. M., Evrard A. E., Richstone D. O., 1996, ApJ, 467, 489

\bibitem[\protect\citename{Dubinski \& Carlberg 1991}]{dub-car91}
Dubinski J., Carlberg R. G., 1991, ApJ, 378, 496

\bibitem[\protect\citename{Dubinski 1994}]{dubinski94}
Dubinski J., 1994, ApJ, 431, 617

\bibitem[\protect\citename{Efstathiou \etal 1988}]
{efstathiou-etal88}
Efstathiou G. P., Frenk C. S., White S. D. M., Davis M., 1988, 
MNRAS, 325, 715

\bibitem[\protect\citename{Eisenstein \& Hut, 1998}]
{eis-hut98}
Eisenstein D. J., Hut P., 1998, ApJ, 498, 137

\bibitem[\protect\citename{Fischer et al. 1997}]{fischer-etal97}
Fischer P., Bernstein G., Rhee G., Tyson J. A., 1997, AJ, 113, 
521

\bibitem[\protect\citename{Fischer \& Tyson 1997}]{fis-tys97}
Fischer P., Tyson J. A., 1997, AJ, 114, 4

\bibitem[\protect\citename{Flin et al. 1995}]
{flin-etal95}
Flin P., Tr\`{e}vese D., Cirimele G., Hickson P., 1995, A\&AS, 
110, 313

\bibitem[\protect\citename{Flin et al. 2000}]
{flin-etal00}
Flin P., Krywult J., Tr\`{e}vese D., Cirimele G., Hickson P., 
A\&ASS, 146, 373 

\bibitem[\protect\citename{Flin, Krywult \& Biernacka 2004}]
{flin-etal04}
Flin P., Krywult J., Biernacka M., 2004, astro-ph/0404182 (FKB)

\bibitem[\protect\citename{Flores \etal 2005}]
{flores-etal05}
Flores R. A., Allgood B., Kravtsov A. V., Primack J. R., 
Buote D. A., Bullock J. S., MNRAS, submitted, astro-ph/0508226

\bibitem[\protect\citename{Floor \etal 2003}]
{floor-etal03}
Floor S. N., Melott A. L., Miller C. J., Bryan G. L., 2003, 
ApJ, 591, 741 

\bibitem[\protect\citename{Floor, Melott \& Motl 2004}]
{floor-etal04}
Floor S. N., Melott A. L., Motl P. M., 2004, ApJ, 611, 153 (FMM)

\bibitem[\protect\citename{Frenk \etal 1985}]{frenk-etal85}
Frenk C. S., White S. D. M., Efstathiou G. P., Davis M., 1985, 
Nature, 317, 595

\bibitem[\protect\citename{Frenk \etal 1988}]{frenk-etal88}
Frenk C. S., White S. D. M., Davis M., Efstathiou G. P., 1988, 
ApJ, 327, 507

\bibitem[\protect\citename{Frenk \etal 1999}]{frenk-etal99}
Frenk C. S. et al., 1999, ApJ, 525, 554

\bibitem[\protect\citename{Gao \etal 2004a}]{gao-etal04a}
Gao L., de Lucia G., White S. D. M., Jenkins A., 2004a, MNRAS, 
352, L1

\bibitem[\protect\citename{Gao \etal 2005b}]{gao-etal04b}
Gao L., White S. D. M., Jenkins A., Stoehr F., Springel V., 
2004b, MNRAS, 355, 819

\bibitem[\protect\citename{Ghigna \etal 1998}]{ghigna-etal98}
Ghigna S., Moore B., Governato F., Lake G., Quinn T., Stadel J., 
1998, MNRAS, 300, 146

\bibitem[\protect\citename{Girardi \etal 1998}]{girardi-etal98}
Girardi M., Borgani S., Giuricin G., Mardirossian F., Mezzetti M., 
1998, ApJ, 506, 45

\bibitem[\protect\citename{Hoekstra 2003}]{hoekstra03}
Hoekstra H., 2003, MNRAS, 339, 1155

\bibitem[\protect\citename{Hoekstra, Yee \& Gladders 2004}]
{hoekstra-etal04}
Hoekstra H., Yee H. K. C., Gladders M. A., 2004, ApJ, 606, 67

\bibitem[\protect\citename{Hobson, Jones \& Lasenby 1999}]
{hobson-etal99}
Hobson M. P., Jones A. W., Lasenby A. N., 1999, MNRAS, 309, 125

\bibitem[\protect\citename{Hopkins, Bahcall \& Bode 2005}]
{hopkins-etal05}
Hopkins P. F., Bahcall N. \& Bode P., 2005, ApJ, 618, 1 

\bibitem[\protect\citename{Jeltema \etal 2005}]{jeltema-etal05}
Jeltema T. E., Canizares C. R., Bautz M. W., Buote D. A., 2005, 
ApJ, 624, 606

\bibitem[\protect\citename{Jing \etal 1995}]{jing-etal95}
Jing Y. P., Mo H. J., B\"{o}rner G., Fang L. Z., 1995, MNRAS, 
276, 417 

\bibitem[\protect\citename{Jing \& Suto 2002}]{jin-sut02}
Jing Y. P., Suto Y, 2002, ApJ, 574, 538

\bibitem[\protect\citename{Kang \etal 2005}]{kang-etal05}
Kang X., Mao S., Gao l., Jing Y. P., 2005, A\&A, 437, 383
 
\bibitem[\protect\citename{Kazantzidis \etal 2004}]
{kazantzidis-etal04}
Kazantzidis S., Kravtsov A. V., Zentner A. R., Allgood B., 
Nagai D., Moore B., 2004, ApJ, 611, L73 

\bibitem[\protect\citename{Kerscher \etal 2001a}]{kerscher-etal01a}
Kerscher M., Mecke K., Schmalzing J., Beisbart C., Buchert T.,
Wagner H., 2001a, A\&A, 373, 1

\bibitem[\protect\citename{Kerscher \etal} 2001b] {kerscher-etal01b}
Kerscher M. et al., 2001b, A\&A, 377, 1

\bibitem[\protect\citename{Kochanek 2002}]{kochanek02}
Kochanek C. S., 2002, in The Shapes of Galaxies and Their Dark 
Matter Halos, ed. Priyamvada Natrajan (World Scientific, Sigapore), 
p. 62, astro-ph/0106495

\bibitem[\protect\citename{Kolb \& Turner} 1990] {kol-tur89}
Kolb E. W., Turner M. S., 1990, Early Universe (Addison-Wesley 
Publishing Company, USA)

\bibitem[\protect\citename{Kolokotronis \etal 2001}]
{kolokotronis-etal97}
Kolokotronis V., Basilakos S., Plionis M., Georgantopoulos I., 
2001, MNRAS, 320, 49

\bibitem[\protect\citename{Libeskind et al. 2005}]{libeskind-etal05}
Libeskind N. I. et al., 2005, MNRAS, 363, 146

\bibitem[\protect\citename{Maccio et al. 2005}]{maccio-etal05}
Maccio A. V., Moore B., Stadel J., Diemand J., 2005, MNRAS, 
submitted, astro-ph/0506125

\bibitem[\protect\citename{Mecke, Buchert \& Wagner 1994}]
{mecke-etal94}
Mecke K. R., Buchert T., Wagner H., 1994, A\&A, 288, 697

\bibitem[\protect\citename{McMillan, Kowalski \& Ulmer 1989}]
{mcmillan-etal89}
McMillan S. L. W., Kowalski M. P., Ulmer M. P., 1989, ApJS, 
70, 723 (MKU)

\bibitem[\protect\citename{Melott, Chambers \& Miller 2001}]
{melott-etal01}
Melott A. L., Chambers S. W., Miller C. J., 2001, ApJ, 559, L75 
(MCM)

\bibitem[\protect\citename{Minkowski 1903}]{minkowski03}
Minkowski H., 1903, Math. Ann., 57, 447

\bibitem[\protect\citename{Motl \etal 2004}]{motl-etal04}
Motl P. M., Burns J. O., Loken C., Norman M. L., Bryan G., 
2004, ApJ, 606, 635 

\bibitem[\protect\citename{Nagai \& Kravtsov 2005}]{nag-kra05}
Nagai D.,  Kravtsov A. V., 2005, ApJ, 618, 557

\bibitem[\protect\citename{Narayan \& Medvedev 2001}]{nar-med01}
Narayan R., Medvedev M. V., 2001, ApJ, 562, L129 

\bibitem[\protect\citename{Natrajan \& Springel 2004}] {nat-vol04}
Natrajan P., Springel V., 2004, ApJ, 617, L13

\bibitem[\protect\citename{Norman \& Bryan 1999}]{nor-bry99}
Norman M. L., Bryan G. L., 1999, ASSL Vol. 240: Numerical 
Astrophysics, eds. S. M. Miyama, K. Tomosaka, T. Hanawa 
(Kluwer Academic Publishers, Boston), p. 19

\bibitem[\protect\citename{Paz \etal. 2005}]{paz-etal05}
Paz D. J., Lambas D. G., Padilla N., Merchin M., 2005, MNRAS, 
submitted, astro-ph/0509062

\bibitem[\protect\citename{Plionis 2002}]{plionis02} 
Plionis M., 2002, ApJ, 572, L67

\bibitem[\protect\citename{Quinn, Salmon \& Zurek 1986}]
{quinn-etal86}
Quinn P. J., Salmon J. K., Zurek W. H., 1986, Nature, 322, 329

\bibitem[\protect\citename{Rhee, van Haarlem \& Katgert 1991}]
{rhee-etal91}
Rhee G. F. R. N., van Haarlem M., Katgert P., 1991, A\&AS, 91, 
513

\bibitem[\protect\citename{Rahman \& Shandarin 2003}]{rah-sh03}
Rahman N., Shandarin S. F., 2003, MNRAS, 343, 933 (RS03)

\bibitem[\protect\citename{Rahman \& Shandarin 2004}]{rah-sh04}
Rahman N., Shandarin S. F., 2004, MNRAS, 354, 235 (RS04)

\bibitem[\protect\citename{Rahman \etal 2004}]{rah-etal04}
Rahman N., Shandarin S. F., Motl P. M., Melott A. L., 2004, 
astro-ph/0405097

\bibitem[\protect\citename{Sarazin 1986}]{sarazin86}
Sarazin C. L., 1988,  X-ray emission from clusters of galaxies, 
Cambridge University Press, Cambridge, New York 
 
\bibitem[\protect\citename{Schmalzing 1999}]{schmalzing99}
Schmalzing J., 1999, Ph.D. Thesis, 
Ludwig Maximilians Universit\"{a}t, M\"{u}nichen, Germany
 
\bibitem[\protect\citename{Schmalzing \etal 1999}]{schmalzing-etal99}
Schmalzing J., Buchert T., Melott A. L., Sahni V.,
Sathyaprakash B. S., Shandarin S. F., ApJ, 526, 568

\bibitem[\protect\citename{Shandarin, Sheth \& Sahni 2003}]
{shandarin-etal04}
Shandarin S. F., Sheth J. V., Sahni V., 2004, MNRAS, 353, 162

\bibitem[\protect\citename{Spergel \etal 2003}]{spergel-etal03}
Spergel D. N. et al., 2003, ApJS, 148, 175

\bibitem[\protect\citename{ Springel, White \& Hernquist 2004}]
{vol-etal04}
Springel V., White S. D. M., Hernquist L., 2004, in Dark Matter 
in Galaxies, IAU Symposium, vol. 220, eds. S. D. Ryder, D. J. 
Pisano, M. A. Walker \& K. C. Freeman, p. 421

\bibitem[\protect\citename{Suwa \etal 2003}]{suwa-etal03}
Suwa T., Habe A., Yoshikawa K., Okamoto T., 2003, ApJ, 588, 17

\bibitem[\protect\citename{Tissera \& Dominguez-Tenreiro 1998}]
{tis-dom98}
Tissera P. B., Dominguez-Tenreiro R., 1998, MNRAS, 297, 177

\bibitem[\protect\citename{Thomas \etal 1998}]{thomas-etal98}
Thomas P. A. et al., 1998, MNRAS, 296, 1061

\bibitem[\protect\citename{Tr\`{e}vese et al. 1992}]{trevese-etal92}
Tr\`{e}vese D., Flin P., Migliori L., Hickson P., Pittella G., 
1992, A\&ASS, 94, 327

\bibitem[\protect\citename{Tr\`{e}vese et al. 1997}]{trevese-etal97}
Tr\`{e}vese D., Cirimele G., Cenci A., Appodia B., 
Flin P., Hickson P., 1997, A\&ASS, 125, 459

\bibitem[\protect\citename{Valdarnini, Ghizzardi \& Bomometto 1999}]
{valdarnini-etal99}
Valdarnini R., Ghizzardi S., Bomometto S., 1999, New Astronomy, 
4, 71

\bibitem[\protect\citename{van den Bosch et al. 2005}]{bosch-etal05} 
van den Bosch F. c., Yang X., Mo H. J., Norberg P., 2005, MNRAS, 
356, 1233

\bibitem[\protect\citename{Voit \etal 1992}]{voit-etal02} 
Voit M G., Bryan G. L., Balogh M. L., Bower R. G., 2002, ApJ, 
576, 601 

\bibitem[\protect\citename{Warren \etal 1992}]{warren-etal92}
Warren M. S., Quinn P. J., Salmon J. K., Zurek W. H., 1992, 
ApJ, 399, 405

\bibitem[\protect\citename{West \& Bothun 1992}]{wes-bot92}
West M. J., Bothun G. D., 1990, ApJ, 350, 36

\bibitem[\protect\citename{Westbury \& Henriksen 1992}]{wes-hen92}
Westbury C. F., Henriksen R. N., 1992, ApJ, 338, 64 

\bibitem[\protect\citename{Zentner et al. 2005}]{zentner-etal05} 
Zentner A. R., Berlind A. A., Bullock J. S., Kravtsov A V., 
Wechsler R. H., 2005, ApJ, 624, 505
\end{thebibliography}
\end{document}